\documentclass[%
nofootinbib,
article,
amsmath,amssymb,
aps,
prd
]{revtex4-2}
\usepackage{color}
\usepackage{graphicx}
\usepackage{dcolumn}
\usepackage{bm}
\usepackage[colorlinks=true,
            linkcolor=blue,
            citecolor=magenta,
            urlcolor=blue]{hyperref}
\usepackage{slashed}
\usepackage{physics}
\usepackage{amsmath}
\usepackage{cancel}
\usepackage{cancel}
\usepackage{soul}
\usepackage{mathrsfs}
\usepackage{dsfont}
\usepackage{tikz}
\usetikzlibrary{trees}
\usetikzlibrary{decorations.pathmorphing}
\usetikzlibrary{decorations.markings}
\tikzset{beamerprimary/.style={structure.fg, thick}}
\tikzset{beamersecondary/.style={structure.bg, thick}}
\tikzset{boson/.style={draw=structure.fg,decorate, decoration={snake}},
    gauge/.style={decorate, decoration={snake} },
    fermion/.style={postaction={decorate},
        decoration={markings,mark=at position .5 with {\arrow{>}}}},
    fermionloop/.style={postaction={decorate},
        decoration={markings,mark=at position .25 with {\arrow{>}}}},
    gluon/.style={decorate,
        decoration={coil,amplitude=4pt, segment length=5pt}},
    scalar/.style={dashed},
    graviton/.style={double}
}
\usepackage{orcidlink}
\newcommand{\shortminus}{\mathbin{\scalebox{0.7}[1]{$-$}}}
\begin{document}


\title{One-Loop Renormalization of Spin-$1/2$ SOPHIES}

\author{I. D\'iaz-Salda\~na~\orcidlink{0009-0004-9658-6700}}\email{isaacdiaz@fisica.uaz.edu.mx}
\affiliation{Unidad Acad\'emica de F\'isica, Universidad Aut\'onoma de Zacatecas, Calzada Solidaridad esquina con Paseo a la Bufa S/N, Zacatecas 98060, Zacatecas, M\'exico}
\author{A. Guti\'errez-Rodr\'iguez\orcidlink{0000-0003-4431-3159}}\email{alexgu@fisica.uaz.edu.mx}
\affiliation{Unidad Acad\'emica de F\'isica, Universidad Aut\'onoma de Zacatecas, Calzada Solidaridad esquina con Paseo a la Bufa S/N, Zacatecas 98060, Zacatecas, M\'exico}

\author{Armando de la C. Rangel-Pantoja~\orcidlink{0009-0004-5315-4086}}\email{adlc.rangelpantoja@ugto.mx}
\affiliation{Departamento de F\'isica, Divisi\'on de Ciencias e Ingenier\'ias, Campus Le\'on, Universidad de
Guanajuato, Loma del Bosque 103, Lomas del Campestre, Le\'on 37150, Guanajuato, M\'exico}
\affiliation{Physics Department, TUM School of Natural Sciences, Technical University of Munich, James-Franck-Strasse 1, 85748, Garching, Germany}
  
 \author{Carlos A. Vaquera-Araujo ~\orcidlink{0000-0001-8578-9263}}\email{vaquera@fisica.ugto.mx}
\affiliation{Secretar\'ia de Ciencia, Humanidades, Tecnolog\'ia e Innovaci\'on, Insurgentes Sur 1582. Colonia Cr\'edito Constructor, Benito Ju\'arez 03940, Ciudad de M\'exico, M\'exico}
\affiliation{Departamento de F\'isica, Divisi\'on de Ciencias e Ingenier\'ias, Campus Le\'on, Universidad de
  Guanajuato, Loma del Bosque 103, Lomas del Campestre, Le\'on 37150, Guanajuato, M\'exico}
\affiliation{Dual CP Institute of High Energy Physics, Colima 28045, Colima, M\'exico}

\date{\today}

\begin{abstract}
In this paper, three formulations of spin-$1/2$ Second-Order Pseudo-Hermitian quantum field theorIES (SOPHIES) are built from scratch from parity eigenstates. Pseudo-Hermiticity allows the unconstrained fields to satisfy either Fermi--Dirac or Bose--Einstein statistics, while self-conjugate Majorana-like fields are only compatible with Fermi--Dirac. The set of independent four dimension renormalizable self-interactions of the three theories are identified, and the one-loop renormalization of the unconstrained fermionic and bosonic theories coupled to an abelian gauge
boson and enhanced with a Pauli term and including self-interactions is studied, together with the one-loop renormalization of the only independent quartic self-interaction of self-conjugate fields. The analysis leads to a rich set of Renormalization Group Equations with deep connections to known theories, like the Dirac formalism and the Klein--Gordon scalar through the fixed points of the Pauli coupling.
\end{abstract}

\maketitle

\section{Introduction}
A free field that satisfies the Dirac equation automatically satisfies the Klein--Gordon (KG) equation. Is it possible to consistently describe the dynamics of a spin-$1/2$ particle by using only the KG equation as the fundamental equation of motion instead of the Dirac one? Recently, there has been progress in that direction in the context of pseudo-Hermitian quantum field theories~\cite{Ferro-Hernandez:2023ymz,Ahluwalia:2023slc,Rangel-Pantoja:2025mhg}. Nevertheless, this idea has been studied since the introduction of the Dirac formalism.
Second-order spin-$1/2$ fermions were considered by Feynman for the first time in~\cite{Feynman:1951gn}, as a trick to more easily evaluate the path integral of a fermion, following a previous idea presented 
by V. Fock~\cite{Fock:1937dy}. Later, the  $V-A$ structure of weak interactions proposed by Feynman and Gell--Mann was motivated by the representation of Fermi particles by two-component Pauli spinors satisfying a second-order differential
equation~\cite{Feynman:1958ty}.
Similar constructions aimed at simplifying calculations with spin-$1/2$ fermions using a second-order formalism can be found in~\cite{Morgan:1995te, Veltman:1997am}. In parallel to those works, where second-order fields were used as a simplifying tool, the search for a physical quantum field theory (QFT) of spin-$1/2$ particles satisfying the KG equation was pursued by the ELKO program (see ~\cite{Ahluwalia:2022ttu} and references therein), where the spin-$1/2$ fields were constructed as eigenstates of the Charge conjugation operator, as opposed to the Parity operator, upon which the Dirac formalism is based. Interestingly, the construction of a consistent free theory of spin-$1/2$ fields satisfying the KG equation was found independently and simultaneously in~\cite{Ferro-Hernandez:2023ymz,Ahluwalia:2023slc}, from two different starting points: Dirac spinors and ELKOs. The crucial observation in both works was the need to relax the Hermiticity of the theory in favor of pseudo-Hermiticity and the recognition that the unconstrained fields have eight real degrees of freedom.  In this work, we construct the second-order spin-$1/2$ fields from first principles by using eigenstates of the parity operator as building blocks, a choice that simplifies the structure of the fields considerably with respect to~\cite{Ahluwalia:2023slc}, and provides a unified language with the Dirac formalism. 

In addition to rendering the canonical quantization of the theory consistent, pseudo-Hermiticity allows the construction of QFTs that circumvent the standard spin-statistics theorem~\cite{LeClair:2007iy,Robinson:2009xm,Ahluwalia:2022zrm,Ahluwalia:2023slc,Rangel-Pantoja:2025mhg,Bai:2025pvb}. Thus, there are two different consistent free Second-Order Pseudo-Hermitian theoriES (SOPHIES) for unconstrained spin-$1/2$ fields, one endowed with the usual Fermi--Dirac statistics, denoted in the following as F-SOPHY; and a second one that satisfies Bose--Einstein statistics, abbreviated as B-SOPHY.  By constraining the fields to be self-conjugate, a third SOPHY emerges, with Majorana-like fields. This theory is only possible for Fermi--Dirac statistics and will be shortened to M-SOPHY. In this work, we study the canonical quantization of M-SOPHY and its symmetries.

Since the SOPHY fields have mass dimension one in four dimensions, they cannot couple directly to the quarks and leptons of the Standard Model (SM) through renormalizable operators. For this reason, second-order fields have naturally been considered as a weakly interacting massive particle (WIMP) dark matter (DM) candidate~\cite{deGracia:2024umr}. However, this is not the only possible application for these fields, as their dimensionality allows them to admit more power-counting renormalizable interactions than their Dirac counterparts. Therefore, SOPHIES are valuable theoretical laboratories for testing new renormalizable interactions. In addition to the interactions generated by minimal coupling  with an Abelian gauge boson, point-like quartic Nambu--Jona--Lasinio-like self-interactions are also dimension-four
and gauge-invariant operators. Moreover, a Pauli-like interaction coupling non-minimally an Abelian gauge field to the SOPHY, introduced in~\cite{Delgado-Acosta:2010ita} to model a fermion with arbitrary gyromagnetic factor, is also dimension four and potentially renormalizable. In this work, we identify the set of independent quartic self-interactions for each of the three spin-$1/2$ SOPHIES and study the one-loop renormalization of the F- and B-SOPHIES coupled to an abelian $U(1)$ gauge boson in arbitrary gauge, enhanced with a Pauli term and quartic self-interactions; and the one-loop renormalization of the single self-interaction of the M-SOPHY.

The paper is organized as follows: In Section~\ref{sec:fields}, the fields are built from scratch. Section~\ref{sec:pH} revisits the canonical quantization of the fields and their symmetries. In Section~\ref{sec:Fierz},
the independent self-interactions for each SOPHY are identified, and in Section~\ref{sec:ren}, the one-loop renormalization of the three spin-$1/2$ SOPHIES is studied. Finally, conclusions and final remarks are presented in Section~\ref{sec:con}.

\section{Fields in the \texorpdfstring{$(1/2,0)\oplus(0,1/2)$}{DiracSpinor} representation of $SO(1,3)^+$}
\label{sec:fields}
In this section, we build from first principles the fields transforming in the $(1/2,0)\oplus(0,1/2)$ representation of the Restricted Lorentz Group  (RLG) $SO(1,3)^+$, using the eigenstates of the parity operator.
\subsection{The \texorpdfstring{$(1/2,0)$}{left} and \texorpdfstring{$(0,1/2)$}{right} representations}
The generators of the RLG satisfy the Lorentz algebra
\begin{gather}
[J^i,J^j] = i\epsilon^{ijk}J^k, \qquad [J^i,K^j] = i\epsilon^{ijk}K^k,\qquad
[K^i,K^j] = -i\epsilon^{ijk}J^k,
\end{gather}
that in terms of  the operators $\mathbf{J}_{\pm}=(\mathbf{J}\pm i \mathbf{K})/2$  splits into two independent $\mathfrak{su}(2)$ algebras satisfying
\begin{gather}
[J^i_\pm,J^j_\pm] = i\epsilon^{ijk}J^k_\pm, \quad [J^i_{+},J^i_{-}] =0.
\end{gather}
Thus, the RLG irreducible representations (irreps) can be labeled with the eigenvalues of the Casimir operators $\mathbf{J}^2_{\pm}=j_\pm(j_\pm+1)$.
The corresponding active proper  orthochronous Lorentz transformation for a field transforming under the $(j_{+},j_{-})$ irrep is
\begin{equation}
S(\boldsymbol{\theta},\boldsymbol{\varphi})=e^{-i(\mathbf{J}\cdot\boldsymbol{\theta}+\mathbf{K}\cdot\boldsymbol{\varphi})}=e^{-i(\mathbf{J}_{+}+\mathbf{J}_{-})\cdot\boldsymbol{\theta}-(\mathbf{J}_{+}-\mathbf{J}_{-})\cdot\boldsymbol{\varphi}},
\end{equation}
with rotation angles $\boldsymbol{\theta}=\theta\mathbf{\hat{m}}$ and rapidities $\boldsymbol{\varphi}=\varphi\mathbf{\hat{n}}$.
In particular, for the \emph{left-handed} $(1/2,0)$ irrep, with $\mathbf{J}_{-}=0$, the Lorentz transformation becomes
\begin{gather}
S_L(\boldsymbol{\theta},\boldsymbol{\varphi})=e^{-i\mathbf{J}^{(1/2)}\cdot\boldsymbol{\theta}-\mathbf{J}^{(1/2)}\cdot\boldsymbol{\varphi}}\equiv R(\boldsymbol{\theta})B_L(\boldsymbol{\varphi}),
\end{gather}
while for the \emph{right-handed} $(0,1/2)$ irrep, with $\mathbf{J}_{+}=0$, we have
\begin{gather}
S_R(\boldsymbol{\theta},\boldsymbol{\varphi})=e^{-i\mathbf{J}^{(1/2)}\cdot\boldsymbol{\theta}+\mathbf{J}^{(1/2)}\cdot\boldsymbol{\varphi}}\equiv R(\boldsymbol{\theta})B_R(\boldsymbol{\varphi}),
\end{gather}
where $\mathbf{J}^{(1/2)}$ are the $SU(2)$ generators for $j=1/2$, written in terms of the Pauli matrices as
\begin{gather}
\mathbf{J}^{(1/2)}
    =
\frac{\boldsymbol{\sigma}}{2}.
\end{gather}
Using the identity
\begin{equation}
[\mathbf{\hat{m}}\cdot\boldsymbol{\sigma}]^2=\mathds{1},
\end{equation}
the explicit form of the rotations is
 \begin{equation}
\begin{split}
R(\boldsymbol{\theta})&=\cos(\theta/2)\mathds{1}-i\sin(\theta/2)\mathbf{\hat{m}}\cdot\boldsymbol{\sigma},
\end{split}
\end{equation}
while under boosts, left- and right-handed Weyl spinors transform according to 
 \begin{equation}
\begin{split}
B_{L,R}(\boldsymbol{\varphi})&=\cosh(\varphi/2)\mathds{1}\mp\sinh(\varphi/2)\mathbf{\hat{n}}\cdot\boldsymbol{\sigma}.
\end{split}
\end{equation}
The group $SU(2)$ is a pseudoreal group and, therefore, there is a matrix $\epsilon$ that relates the rotation matrix to its conjugate as
$
R(\boldsymbol{\theta})^*=\epsilon^{-1} R(\boldsymbol{\theta})\epsilon.
$
The explicit form of this matrix is
\begin{align}
    \epsilon=&
     i\sigma^2
     =
     \begin{pmatrix}
         0 & 1  \\
         -1 & 0 \\
     \end{pmatrix},
 \end{align}
 satisfying $\epsilon=-\epsilon^{-1}=-\epsilon^\dagger=-\epsilon^{T}=\epsilon^{*}$, and 
 \begin{gather}
[\mathbf{\widehat{m}}\cdot \boldsymbol{\sigma}]^{*} 
     =\epsilon [\mathbf{\widehat{m}}\cdot \boldsymbol{\sigma} ] \epsilon.\label{conjeps}
 \end{gather}
We can write the rapidity in terms of energy and momentum according to the relations $\cosh\varphi=\gamma=\frac{\omega_{\mathbf{p}}}{m}$, $\sinh\varphi=\gamma\beta=\frac{|\mathbf{p}|}{m}$, with $\omega_{\mathbf{p}}=+\sqrt{\mathbf{p^2}+m^2}$, thus the boosts are explicitly
\begin{equation}\label{BoostLR}
\begin{split}
B_{L,R}(\mathbf{p})
&=\frac{(\omega_\mathbf{p}+m)\mathds{1}\mp\mathbf{p}\cdot\boldsymbol{\sigma}}{\sqrt{2m(\omega_\mathbf{p}+m)}},
\end{split}
\end{equation}
and satisfy \begin{equation}
B^\dagger_{L,R}(\mathbf{p})=B_{L,R}(\mathbf{p}),\qquad  B_{R,L}(\mathbf{p})B_{L,R}(\mathbf{p})=\mathds{1}.
\end{equation}
The squared left and right boost operators can be written as
\begin{align}
    B_{L}(\mathbf{p})^2=\frac{p\cdot\sigma}{m}, \quad B_{R}(\mathbf{p})^2=\frac{p\cdot\overline{\sigma}}{m},
\end{align}
with $\sigma^\mu=(\mathds{1},\boldsymbol{\sigma})$ and $\overline{\sigma}^\mu=(\mathds{1},-\boldsymbol{\sigma})$.   
The product $(p\cdot\sigma)(p\cdot\overline{\sigma})$ can be evaluated explicitly and reduces to
\begin{equation}
    (p\cdot\sigma)(p\cdot\overline{\sigma})
    =p^{\mu}p_{\mu}=p^2,
\end{equation}
for an arbitrary off-shell momentum $p^{\mu}$, and on-shell to $(p\cdot\sigma)(p\cdot\overline{\sigma})
    =m^2$.
In terms of $p\cdot\sigma$ and $p\cdot\overline{\sigma}$, we can write the left and right boost operators $B_{L}(\mathbf{p})$, $B_{R}(\mathbf{p})$ compactly as
\begin{align}
    B_{L}(\mathbf{p})
    =\sqrt{\frac{p\cdot\sigma}{m}},\quad  B_{R}(\mathbf{p})
    =\sqrt{\frac{p\cdot\overline{\sigma}}{m}}.
\end{align}
The Lorentz generators for the $(1/2,0)$ and $(0,1/2)$ irreps become, respectively 
\begin{align}
M_{L}^{\mu\nu}&=\frac{i}{4}\left( \sigma^{\mu}\overline{\sigma}^{\nu}- \sigma^{\nu}\overline{\sigma}^{\mu} \right),\\
M_{R}^{\mu\nu}&=\frac{i}{4}\left( \overline{\sigma}^{\mu}\sigma^{\nu}-\overline{\sigma}^{\nu}\sigma^{\mu}  \right),
\end{align}
and the Lorentz generators of the $(1/2,0)\oplus(0,1/2)$ representation in the chiral basis are given by 
\begin{equation}
M^{\mu\nu}=\begin{pmatrix}
    M_{L}^{\mu\nu} & 0\\
    0 & M_{R}^{\mu\nu}  
\end{pmatrix}.
\end{equation}
In this representation, \emph{Parity} is defined as the transformation $\Pi^{-1}\mathbf{J}_{+}\Pi=\mathbf{J}_{-}$, $\Pi^{-1}\mathbf{J}_{-}\Pi=\mathbf{J}_{+}$, and so it
exchanges $(j_{+},j_{-})\leftrightarrow (j_{-},j_{+})$ irreps.
In this basis, the boost operator for the direct sum representation is 
\begin{equation}\label{Boost1}
B(\mathbf{p})=\begin{pmatrix}
    B_{L}(\mathbf{p}) & 0\\
    0 & B_{R}(\mathbf{p})  
\end{pmatrix},
\end{equation}
while the parity operator reads
\begin{equation}
\Pi=\begin{pmatrix}
0 & 1\\
1 & 0
\end{pmatrix},
\end{equation}
with $\Pi=\Pi^{-1}=\Pi^\dagger=\Pi^T=\Pi^*$.
As shown below, this operator is the key to building the most general field transforming under the $(1/2,0)\oplus(0,1/2)$ representation of the RLG and satisfying the KG equation. 

\subsection{Parity Eigenvectors, Charge Conjugation and Chiral Operator}

 As a first step towards the construction of the most general spin-$1/2$ field satisfying the KG equation and transforming under the $(1/2,0)\oplus(0,1/2)$ rep of the RLG, we identify the eigenvectors of the parity operator $\Pi$, which are four-component column spinors, denoted here as $u^s_{\mathbf{0}}$ and $v^s_{\mathbf{0}}$
\begin{equation}
    \Pi u^s_{\mathbf{0}}=u^s_{\mathbf{0}},\qquad \Pi v^s_{\mathbf{0}}=-v^s_{\mathbf{0}}.
\end{equation}
These spinors can be written as two-component blocks
\begin{align}\label{ParEig}
    u_{\mathbf{0}}^{s}
    \equiv
    \sqrt{m}\begin{pmatrix}
        \phi^{s}
        \\
        \phi^{s}
    \end{pmatrix} \quad v_{\mathbf{0}}^{s}
    \equiv
    \sqrt{m}\begin{pmatrix}
        \chi^{s}
        \\
        -\chi^{s}
    \end{pmatrix},
\end{align}
where $\phi^{s}$, $s=\pm 1/2$, are Pauli spinors describing spin orientation
\begin{align}
    \phi^{1/2}
    =
    \begin{pmatrix}
        1\\0
    \end{pmatrix},\quad \phi^{-1/2}=
    \begin{pmatrix}
        0\\1
    \end{pmatrix},
\end{align}
satisfying the following orthonormal and completeness relations 
\begin{align}
    \phi^{r\dagger}\phi^{s}= \delta^{rs},\quad \sum_{s}\phi^{s}_{a}\phi^{s\dagger}_{b}=\delta_{ab}.
\end{align}
Their charge-conjugate versions are defined as $\chi^{s}\equiv-\epsilon\phi^{s\,*}$, and read in components 
\begin{gather}
    \chi^{1/2}
    =
    \begin{pmatrix}
        0\\1
    \end{pmatrix}
    =
    \phi^{-1/2},
    \quad
    \chi^{-1/2}
    =
    \begin{pmatrix}
        -1\\0
    \end{pmatrix}
    =
    -\phi^{1/2},
\end{gather}
or equivalently 
\begin{equation}\label{chixi}
\chi^s=2s\phi^{-s}.
\end{equation}

Boosting the parity operator, we get
\begin{equation}
\label{BoostingParity}
\begin{split}
    B(\mathbf{p}) \Pi B^{-1}(\mathbf{p})
    =&
    \frac{\gamma^{\mu}
    p_{\mu}}{m}
    \equiv
    \frac{\slashed{p}}{m},
\end{split}
\end{equation}
where $\gamma^\mu$ are the Dirac matrices in the chiral basis
\begin{equation}
\label{SmunuCovariant}
    \gamma^{\mu}
    \equiv
    \begin{pmatrix}
        0 &\sigma^{\mu}\\
         \overline{\sigma}^{\mu} & 0
    \end{pmatrix}.  
\end{equation}
These matrices satisfy the Clifford algebra $\{\gamma^\mu,\gamma^\nu\}=2g^{\mu\nu}\mathds{1}$ and $(\gamma^\mu)^\dagger=\gamma^0\gamma^\mu\gamma^0$. In terms of the Dirac matrices, the Lorentz generators for the $(1/2,0)\oplus(0,1/2)$ representation become
\begin{align}
M^{\mu\nu}&=\frac{i}{4}[\gamma^{\mu},\gamma^{\nu}].
\end{align}
Notice that on this basis, the parity operator is $\Pi=\gamma^{0}$.
The Dirac spinors $u_{\mathbf{p}}^{s}$ and $v_{\mathbf{p}}^{s}$ with arbitrary momentum $\mathbf{p}$ can be obtained by boosting the parity eigenvectors in Eq.~\eqref{ParEig}
\begin{gather}
    u_{\mathbf{p}}^{s}
    =
    B(\mathbf{p}) u_{\mathbf{0}}^{s}
    =
    \sqrt{m}
    \begin{pmatrix}
        B_{L}(\mathbf{p}) \phi^{s} \\
        B_{R}(\mathbf{p}) \phi^{s}
    \end{pmatrix}
    =
\begin{pmatrix}
        \sqrt{p\cdot\sigma} \phi^{s} \\
        \sqrt{p\cdot\overline{\sigma} }\phi^{s}
    \end{pmatrix},
\\
    v_{\mathbf{p}}^{s}
    =
    B(\mathbf{p}) v_{\mathbf{0}}^{s}
    =
    \sqrt{m}
    \begin{pmatrix}
        B_{L}(\mathbf{p}) \chi^{s} \\
        \shortminus B_{R}(\mathbf{p}) \chi^{s}
    \end{pmatrix}
    =
\begin{pmatrix}
        \sqrt{p\cdot\sigma} \chi^{s} \\
        \shortminus\sqrt{p\cdot\overline{\sigma}} \chi^{s}
    \end{pmatrix}
.
\end{gather}
Lorentz scalars can be built with the aid of the Dirac adjoints, which are defined as
\begin{align}
\overline{u}_{\mathbf{p}}^{s}=& u_{\mathbf{p}}^{s\,\dagger}\gamma^{0}
    =
    \begin{pmatrix}
        \phi^{s\,\dagger}\sqrt{p\cdot\overline{\sigma}}  &
        \phi^{s\,\dagger}\sqrt{p\cdot\sigma}  
    \end{pmatrix} ,\\
    \overline{v}_{\mathbf{p}}^{s}=& v_{\mathbf{p}}^{s\,\dagger}\gamma^{0}
    =
    \begin{pmatrix}
        \shortminus\chi^{s\,\dagger}\sqrt{p\cdot\overline{\sigma} }&
       \chi^{s\,\dagger}\sqrt{ p\cdot\sigma }
    \end{pmatrix}.
\end{align}

The charge-conjugate boosted parity eigenvectors $u_{\mathbf{p}}^{s\,c}$ and $v_{\mathbf{p}}^{s\,c}$ are defined as
\begin{align}\label{charcon}
    u_{\mathbf{p}}^{s\,c}
    \equiv \,
    \mathcal{C}
    \overline{u}_{\mathbf{p}}^{s\,T},\quad v_{\mathbf{p}}^{s\,c}\equiv\,\mathcal{C}\overline{v}_{\mathbf{p}}^{s\,T},
\end{align}
with 
\begin{align}
    \mathcal{C}=&
    \begin{pmatrix}
        \shortminus\epsilon & 0 \\
         0 & \epsilon
    \end{pmatrix},
\end{align}
satisfying $\mathcal{C}^{-1}=\mathcal{C}^\dagger=\mathcal{C}^{T}=-\mathcal{C}^{*}=-\mathcal{C}$ and $\{\mathcal{C},\Pi\}=0$.
Explicitly
\begin{align}
    u_{\mathbf{p}}^{s\,c}= 
    \sqrt{m}
    \begin{pmatrix}
       \shortminus\epsilon B_{R}^{*}(\mathbf{p})\phi^{s\,*}
        \\
       \epsilon B_{L}^{*}(\mathbf{p})\phi^{s\,*}
    \end{pmatrix}, \quad v_{\mathbf{p}}^{s\,c}= 
    \sqrt{m}
    \begin{pmatrix}
        \epsilon B_{R}^{*}(\mathbf{p})\chi^{s\,*}
        \\
       \epsilon B_{L}^{*}(\mathbf{p})\chi^{s\,*}
    \end{pmatrix}.
\end{align}
By combining Eq.~\eqref{conjeps} and Eq.~\eqref{BoostLR}, it can be shown that
\begin{align}
    \epsilon B^{*}_{R}(\mathbf{p})\epsilon=-B_{L}(\mathbf{p}),\quad \epsilon B^{*}_{L}(\mathbf{p})\epsilon=-B_{R}(\mathbf{p}),
\end{align}
and therefore, the charge-conjugate boosted parity eigenvectors become
\begin{align}
    &u_{\mathbf{p}}^{s\,c}
    =
    \sqrt{m}
    \begin{pmatrix}
        B_{L}(\mathbf{p})\chi^{s}
        \\
        \shortminus B_{R}(\mathbf{p})\chi^{s}
    \end{pmatrix}
    =
\begin{pmatrix}
        \sqrt{p\cdot\sigma} \chi^{s} \\
        \shortminus\sqrt{p\cdot\overline{\sigma}} \chi^{s}
    \end{pmatrix}=v_{\mathbf{p}}^{s},
    \\
    &v_{\mathbf{p}}^{s\,c}
    =
    \sqrt{m}
    \begin{pmatrix}
        B_{L}(\mathbf{p})\phi^{s}
        \\
        B_{R}(\mathbf{p})\phi^{s}
    \end{pmatrix}
=\begin{pmatrix}
        \sqrt{p\cdot\sigma} \phi^{s} \\
\sqrt{p\cdot\overline{\sigma}} \phi^{s}
    \end{pmatrix}=u_{\mathbf{p}}^{s}.
\end{align}

The chirality operator is defined as the Casimir operator
\begin{equation}
\gamma^5\equiv i\widetilde{M}^{\mu\nu}M_{\mu\nu},
\end{equation}
where $\widetilde{M}^{\mu\nu}=(1/2)\epsilon^{\mu\nu\rho\sigma}M_{\rho\sigma}$ are the dual Lorentz generators in the $(1/2,0)\oplus(0,1/2)$ representation, with $\epsilon^{\mu\nu\rho\sigma}$ being the totally antisymmetric Levi-Civita symbol in Minkowski space-time with $\epsilon^{0123}\equiv1$. In the chiral basis, the chirality operator becomes
\begin{equation}
\gamma^5=\begin{pmatrix}
\shortminus 1 & 0\\
0  & 1
\end{pmatrix}, 
\end{equation}
and satisfies $\{\gamma^{\mu},\gamma^5\}=0$, $(\gamma^5)^\dagger=\gamma^5$, $(\gamma^5)^2=1$, $[\gamma^5,\mathcal{C}]=0$. Operating $\gamma^5$ on a boosted parity eigenvector yields another boosted parity eigenvector
\begin{equation}
\gamma^5u_{\mathbf{p}}^{s}=2sv_{\mathbf{p}}^{\shortminus s},\qquad \gamma^5v_{\mathbf{p}}^{s}=-2su_{\mathbf{p}}^{\shortminus s}.
\end{equation}
Dirac spinors obey the following orthogonality relations
\begin{gather}
\bar{u}_{\mathbf{p}}^{r} u_{\mathbf{p}}^{s}=2m\delta^{rs},\quad
\bar{v}_{\mathbf{p}}^{r} v_{\mathbf{p}}^{s}=-2m\delta^{rs},
\\
\bar{u}_{\mathbf{p}}^{r} v_{\mathbf{p}}^{s}=0,
\quad
\bar{v}_{\mathbf{p}}^{r} u_{\mathbf{p}}^{s}=0,
\\
\bar{u}_{\mathbf{p}}^{r}\gamma^5 u_{\mathbf{p}}^{s}=0,
\quad
\bar{v}_{\mathbf{p}}^{r} \gamma^{5}v_{\mathbf{p}}^{s}=0,
\\
\bar{u}_{\mathbf{p}}^{r}\gamma^5 v_{\mathbf{p}}^{s}=-4sm\delta^{r,\shortminus s},
\quad
\bar{v}_{\mathbf{p}}^{r}\gamma^5 u_{\mathbf{p}}^{s}=-4sm\,\delta^{r,\shortminus s},
\end{gather}
and the outer products
\begin{gather}
\begin{split}
\sum_{s}
u_{\mathbf{p}a}^{s}
\bar{u}_{\mathbf{p}b}^{s}
=-\sum_{s}\gamma^5v_{\mathbf{p}a}^{s}
\bar{v}_{\mathbf{p}b}^{s}\gamma^5
=&
\left(
\slashed{p}+m\right)_{ab},
\end{split}\\
\begin{split}
\sum_{s}
v_{\mathbf{p}a}^{s}
\bar{v}_{\mathbf{p}b}^{s}
=&-\sum_{s}
\gamma^5u_{\mathbf{p}a}^{s}
\bar{u}_{\mathbf{p}b}^{s}\gamma^5=
\left(
\slashed{p}-m\right)_{ab}.
\end{split}
\end{gather}
By construction, the boosted parity eigenvectors $u_{\mathbf{p}}^{s},v_{\mathbf{p}}^{s}$ and $\gamma^5u_{\mathbf{p}}^{s},\gamma^5v_{\mathbf{p}}^{s}$ satisfy on-shell the equations,
\begin{align}
\label{sigmaeq1}
    \left(
    \slashed{p}
    -m
    \right)
    u_{\mathbf{p}}^{s}
    =&0,
    &
    \left(
    \slashed{p}
    -m
    \right)
    \gamma^5v_{\mathbf{p}}^{s}
    =&0,
    \\
    \label{sigmaeq2}
    \left(\slashed{p}
    +m
    \right)
    v_{\mathbf{p}}^{s}
    =&0,
    &
    \left(
    \slashed{p}
    +m
    \right)
    \gamma^5u_{\mathbf{p}}^{s}
    =&0.
\end{align}
Besides, for an off-shell arbitrary momentum $p^\mu$ we have
\begin{equation}
\begin{split}
(\slashed{p}
    +m)(\slashed{p}
    -m)
=(p^2-m^2)\mathds{1}.
\end{split}
\end{equation} 
Thus, the boosted parity eigenspinors $u_{\mathbf{p}}^{s}$, $v_{\mathbf{p}}^{s}$, $\gamma^5u_{\mathbf{p}}^{s}$, and $\gamma^5v_{\mathbf{p}}^{s}$ are all solutions to the on-shell condition $(p^2-m^2)w_{\mathbf{p}}^{s}=0$, and therefore, the most general field transforming under the $(1/2,0)\oplus(0,1/2)$ representation of the RLG that satisfies the KG equation can be written as
\begin{align}
\psi(x)=&\int \frac{d^3\mathbf{p}}{(2\pi)^{3}2\sqrt{m\omega_{\mathbf{p}}}}\sum_s \bigg\{\left(u^{s}_{\mathbf{p}}a^{1 s}_{\mathbf{p}}+\gamma^5u^{s}_{\mathbf{p}}a^{2 s}_{\mathbf{p}}\right)e^{-ip\cdot x}+\left(v^{s}_{\mathbf{p}}b^{1 s\dagger}_{\mathbf{p}}+\gamma^5v^{s}_{\mathbf{p}}b^{2 s\dagger}_{\mathbf{p}}\right)e^{ip\cdot x}\bigg\}.\label{field_expansion_1}
\end{align}
This can be viewed as a superposition of two independent fields satisfying the Dirac equation. Indeed, an unconstrained four-component complex spinor satisfying the KG equation contains eight degrees of freedom, while each Dirac field carries only four~\cite{CufaroPetroni:1985tu}.
\section{Pseudo-Hermiticity and Quantization}\label{sec:pH}
\subsection{Pseudo-Hermitian quantum field theories}
The quantization of the field in Eq.~\eqref{field_expansion_1} can be successfully performed in the framework of a pseudo-Hermitian quantum field theory, where the Lagrangian density  obeys the relation
\begin{equation}
\mathcal{L}^{\#}\equiv \eta^{-1}\mathcal{L}^\dagger\eta=\mathcal{L},
\label{pseudo-Hermitian}
\end{equation}
mediated by a specific operator $\eta$ in Fock space. Although the relaxation of Hermiticity was originally investigated within quantum mechanics~\cite{Mostafazadeh:2001jk, Mostafazadeh:2008pw} as a generalization of $PT$-symmetric systems~\cite{Bender:1998ke}, recent developments have successfully extended these concepts to consistent quantum field theories~\cite{Sablevice:2023odu}. Crucially, despite the non-Hermitian nature of the theory, such theories can ensure a real energy spectrum and unitary time evolution, provided that the inner product is redefined with respect to the $\eta$ operator.

In this context, the dynamics of the $\psi$ field in Eq.~\eqref{field_expansion_1} is governed by the Lagrangian
\begin{equation}\label{Lag1}
\begin{split}
\mathcal{L}= \partial^{\mu}\widehat{\psi}\partial_\mu\psi-m^2\widehat{\psi}\psi ,
\end{split}
\end{equation}
where $\widehat{\psi}$ denotes the modified dual field required to preserve pseudo-Hermiticity. This is related to the standard Dirac adjoint $\bar{\psi}=\psi^{\dagger}\gamma^{0}$, via the operator $\eta$ as follows
\begin{equation}
\widehat\psi=\eta^{-1}\bar\psi\eta.\label{dual}
\end{equation} 
The canonical quantization of two versions of this theory have been presented in~\cite{Ferro-Hernandez:2023ymz,Rangel-Pantoja:2025mhg}, and independently a closely related formalism has been developed in~\cite{Ahluwalia:2023slc} from the perspective of ELKO fields.
The mode expansion for the dual field  $\widehat\psi$ studied in~\cite{Ferro-Hernandez:2023ymz,Rangel-Pantoja:2025mhg} is 
\begin{align}
\widehat{\psi}(x)=&\int \frac{d^3\mathbf{p}}{(2\pi)^{3}2\sqrt{m\omega_{\mathbf{p}}}}\sum_s \bigg\{\left(\bar{u}^{s}_{\mathbf{p}}a^{1 s\dagger}_{\mathbf{p}}+\bar{u}^{s}_{\mathbf{p}}\gamma^5a^{2 s\dagger}_{\mathbf{p}}\right)e^{ip\cdot x}+\varsigma\left(\bar{v}^{s}_{\mathbf{p}}b^{1 s}_{\mathbf{p}}+\bar{v}^{s}_{\mathbf{p}}\gamma^5b^{2 s\dagger}_{\mathbf{p}}\right)e^{-ip\cdot x}\bigg\},\label{field_expansion_2}
\end{align}
with $\omega_{\mathbf{p}}=+\sqrt{|\mathbf{p}|^2+m^2}$, $p^\mu=(\omega_{\mathbf{p}},\mathbf{p})$ and $\varsigma=\pm 1$. In the above expression and what follows,  $\varsigma=+1$ stands for F-SOPHY and $\varsigma=-1$ stands for B-SOPHY.
The explicit form of $\eta$ is found to be
\begin{equation}
\eta=\exp\left[i\pi\int\frac{d^3\mathbf{p}}{(2\pi)^{3}}\sum_s \left(a^{2 s\dagger}_{\mathbf{p}}a^{2 s}_{\mathbf{p}}+b^{\varepsilon s\dagger}_{\mathbf{p}}b^{\varepsilon s}_{\mathbf{p}}\right)\right],
\end{equation}
where
\begin{equation}
 \varepsilon=\frac{3+\varsigma}{2}.   
\end{equation}
In both cases $\eta$ is Hermitian and unitary ($\eta=\eta^{-1}=\eta^\dagger$ and therefore $\eta^2=1$). Finally, it is straightforward to verify that this construction ensures the condition $\eta^{-1}(\widehat\psi\psi)^\dagger\eta=\widehat\psi\psi$, thus satisfying the pseudo-Hermiticity requirement of Eq.~\eqref{Lag1} consistent with the definition in Eq.~\eqref{pseudo-Hermitian}.

A self-conjugate field analogous to a Majorana field is possible only for Fermi-Dirac statistics. The mode expansions for such a field and its modified dual are given by 
\begin{align}
\psi_M(x)=&\int \frac{d^3\mathbf{p}}{(2\pi)^{3}2\sqrt{m\omega_{\mathbf{p}}}}\sum_s \bigg\{\left(u^{s}_{\mathbf{p}}a^{1 s}_{\mathbf{p}}+\gamma^5u^{s}_{\mathbf{p}}a^{2 s}_{\mathbf{p}}\right)e^{-ip\cdot x}+\left(v^{s}_{\mathbf{p}}a^{1 s\dagger}_{\mathbf{p}}+\gamma^5v^{s}_{\mathbf{p}}a^{2 s\dagger}_{\mathbf{p}}\right)e^{ip\cdot x}\bigg\},\label{field_expansion_3}\\
\widehat{\psi}_M(x)=&\int \frac{d^3\mathbf{p}}{(2\pi)^{3}2\sqrt{m\omega_{\mathbf{p}}}}\sum_s \bigg\{\left(\bar{u}^{s}_{\mathbf{p}}a^{1 s\dagger}_{\mathbf{p}}+\bar{u}^{s}_{\mathbf{p}}\gamma^5a^{2 s\dagger}_{\mathbf{p}}\right)e^{ip\cdot x}+\left(\bar{v}^{s}_{\mathbf{p}}a^{1 s}_{\mathbf{p}}+\bar{v}^{s}_{\mathbf{p}}\gamma^5a^{2 s\dagger}_{\mathbf{p}}\right)e^{-ip\cdot x}\bigg\}.\label{field_expansion_4}
\end{align}
This field satisfies
\begin{equation}
\psi_M^c(x)=\psi_M(x),\qquad \psi_M^c(x)\equiv \mathcal{C}\widehat{\psi}_M^T(x),
\end{equation}
with 
\begin{equation}
\widehat{\psi}_M\equiv\eta_M^{-1}\overline{\psi}_M\eta_M=\psi^T_M\mathcal{C},\qquad\eta_M=\exp\left[i\pi\int\frac{d^3\mathbf{p}}{(2\pi)^{3}}\sum_s a^{2 s\dagger}_{\mathbf{p}}a^{2 s}_{\mathbf{p}}\right],
\end{equation}
and is described by the pseudo-Hermitian free Lagrangian density
\begin{equation}
\begin{split}
\mathcal{L}_M= \frac{1}{2}\partial^{\mu}\widehat{\psi}_M\partial_\mu\psi_M-\frac{m^2}{2}\widehat{\psi}_M\psi_M=\frac{1}{2}\partial^{\mu}\psi^T_M\mathcal{C}\partial_{\mu}\psi_M-\frac{m^2}{2}\psi^T_M\mathcal{C}\psi_M.
\end{split}\label{self_conjugate}
\end{equation}

\subsection{Canonical Quantization}
We now summarize the main results of the canonical quantization of the SOPHIES described by the Lagrangian (\ref{Lag1}) and studied in~\cite{Ferro-Hernandez:2023ymz,Rangel-Pantoja:2025mhg}. In standard Hermitian quantum field theories, the quantization procedure is rigidly constrained by the Pauli Spin-Statistics Theorem, which dictates that integer spin fields must be quantized using commutators (Bose-Einstein statistics), while half-integer spin fields require anticommutators (Fermi--Dirac statistics) to ensure causality and the positivity of the energy spectrum. However, the framework of pseudo-Hermitian quantum field theory offers a richer landscape. As discussed in~\cite{Bai:2025pvb}, the introduction of the operator $\eta$ and the subsequent redefinition of the dual space relax the stringent conditions of the standard Spin-Statistics connection. Thus, the canonical quantization of the theory described by~\eqref{Lag1} can be formulated in a unified manner for both fermionic and bosonic sectors. We define the graded bracket $[\cdot,\cdot]_{\varsigma}$ so that the subscript $\varsigma=-1$ denotes the standard commutator (B-SOPHY) and the subscript $\varsigma=+1$ denotes the anticommutator (F-SOPHY). The equal-time quantization conditions for the fields and their conjugate momenta are prescribed by
\begin{equation}\label{can_1}
\begin{split}
\left[\psi_{\alpha}(\mathbf{x},t),\pi_\psi{}_{\beta}(\mathbf{x}',t)\right]_\varsigma &=-\varsigma \left[\widehat{\psi}_{\alpha}(\mathbf{x},t),\pi_{\widehat{\psi}}{}_{\beta}(\mathbf{x}',t)\right]_\varsigma \\
&= i\delta_{\alpha\beta}\delta^{(3)}(\mathbf{x}-\mathbf{x}'),
\end{split}
\end{equation}
where $\alpha,\beta=1,\ldots,4$. The canonical momenta associated with the field $\psi$ and its dual $\widehat{\psi}$ are defined, as usual, by
\begin{equation}
\pi_\psi=\frac{\partial\mathcal{L}}{\partial \dot{\psi}}=\dot{\widehat{\psi}},\qquad\qquad \pi_{\widehat{\psi}}=\frac{\partial\mathcal{L}}{\partial \dot{\widehat{\psi}}}=\dot{\psi},
\end{equation}
where for fermions, it is understood that the derivative with respect to $\dot{\psi}$ is taken from the right, and the derivative with respect to $\dot{\widehat{\psi}}$ from the left.
Implementing the condition \eqref{can_1} leads to the following algebra for the creation and annihilation operators in momentum space:
\begin{equation}
\begin{split}
\left[a^{j s}_{\mathbf{p}},a^{k r\dagger}_{\mathbf{p}'}\right]_\varsigma &=(2\pi)^3\delta^{j k}\delta^{r s}\delta^{(3)}(\mathbf{p}-\mathbf{p}'),\\
\left[b^{j s}_{\mathbf{p}},b^{k r\dagger}_{\mathbf{p}'}\right]_\varsigma &=(2\pi)^3\delta^{j k}\delta^{r s}\delta^{(3)}(\mathbf{p}-\mathbf{p}'),
\end{split}
\label{canonicalrelations}
\end{equation}
with all other mixed brackets vanishing. It is important to emphasize that the  operator $\eta$, implicitly contained within the definition of the dual field $\widehat{\psi}$, is essential for the consistency of these relations. Specifically, $\eta$ is responsible for correcting the sign of the brackets for the $a^{2}$ and $b^{2}\,(b^{1})$ sectors for fermions (bosons), which would otherwise differ from the standard form if the conventional Dirac adjoint were employed.

Furthermore, the quantized theory exhibits a consistent microcausality structure. The equal-time graded brackets of the field and the dual field with themselves vanish identically,
\begin{equation}
    \left[\psi_\alpha(x), \psi_\beta(y)\right]_\varsigma = \left[\widehat{\psi}_\alpha(x), \widehat{\psi}_\beta(y)\right]_\varsigma = 0,
\end{equation}
whereas the mixed bracket between the field and its dual recovers the standard invariant function, characteristic of relativistic field theories
\begin{equation}\label{micro}
 \begin{split}
\left[\psi_\alpha(x),\widehat{\psi}_\beta(y)\right]_\varsigma &=\delta_{\alpha\beta}\Delta(x-y)\\
&=\delta_{\alpha\beta}\int  \frac{d^3\mathbf{p}  }{(2\pi)^{3}2{\omega_{\mathbf{p}}}}\bigg\{ e^{-ip\cdot
   (x- y)}-e^{ip\cdot
   (x- y)}  \bigg\},
    \end{split}
\end{equation}
where $\Delta(x-y)$ represents the Schwinger function.

The dynamical quantities of the system, namely the Hamiltonian and the momentum operators, are given by
\begin{equation}\label{ham}
\begin{split}
H&=:\int d^3\mathbf{x}\bigg(\dot{\widehat{\psi}}\dot{\psi}+\nabla\widehat\psi\cdot\nabla\psi+m^2\widehat\psi\psi\bigg):,\\
\mathbf{P}&=:-\int d^3\mathbf{x}\bigg(\dot{\widehat{\psi}}\nabla\psi+\nabla\widehat{\psi}\dot{ \psi}\bigg):,
\end{split}
\end{equation}
where $:\ :$ stands for normal ordering of the operators; this handles the operator reordering according to the chosen statistics (subtracting the vacuum divergence appropriately for either commutators or anticommutators). Substituting the mode expansions into these expressions yields the following four-momentum operator $P^{\mu}=(H,\mathbf{P})$ 
\begin{equation}
\begin{split}
    P^{\mu}&= :\int  \frac{d^3\mathbf{p}}{(2\pi)^{3}}p^{\mu}\sum_{j,s}\left(a^{j s\dagger}_{\mathbf{p}}a^{j s}_{\mathbf{p}}-\varsigma b^{j s}_{\mathbf{p}}b^{j s\dagger}_{\mathbf{p}}\right):\\
    &=\int  \frac{d^3\mathbf{p}}{(2\pi)^{3}}p^{\mu}\sum_{j,s}\left(a^{j s\dagger}_{\mathbf{p}}a^{j s}_{\mathbf{p}}+ b^{j s\dagger}_{\mathbf{p}}b^{j s}_{\mathbf{p}}\right).
\end{split}
\end{equation}
Consequently, the energy spectrum is real and bounded from below, rendering the canonical quantization consistent. It is worth noting that the free Hamiltonian is explicitly Hermitian when expressed in terms of the momentum-space operators.

The Feynman propagator for the free pseudo-Hermitian theory is defined via the vacuum expectation value of the time-ordered product of the field and its dual. Taking into account the graded algebra introduced previously, the time-ordering operator $T$ incorporates a sign $\pm$ upon field exchange, yielding
\begin{equation}
\begin{split}
&S^F_{\alpha\beta}(x-y)=\bra{0}T\left[\psi_\alpha(x)\widehat{\psi}_\beta(y)\right]\ket{0}\\
&\qquad=\bra{0}\left[\theta(x^0-y^0)\psi_\alpha(x)\widehat{\psi}_\beta(y) -\varsigma\, \theta(y^0-x^0)\widehat{\psi}_\beta(y)\psi_\alpha(x)\right]\ket{0}\\
&\qquad=\int\frac{d^4p}{(2\pi)^4}\left[\frac{i\delta_{\alpha\beta}}{p^2-m^2+i\epsilon}\right]e^{-i p\cdot(x-y)}.
\end{split}
\end{equation}
Remarkably, regardless of whether the fields obey Bose--Einstein or Fermi--Dirac statistics, the resulting expression identically reproduces the standard  propagator expected for a field governed by the KG equation.

Analogous results are obtained for the M-SOPHY described by Eq.~\eqref{self_conjugate}. In this case, canonical quantization follows from
\begin{equation}\label{can_2}
\begin{split}
\left[\psi_{M\alpha}(\mathbf{x},t),\dot{\widehat{\psi}}_{M\beta}(\mathbf{x}',t)\right]_{+} &= i\delta_{\alpha\beta}\delta^{(3)}(\mathbf{x}-\mathbf{x}'),
\end{split}
\end{equation}
where we have denoted the anticommutator as $[\cdots]_{\varsigma=+1}\equiv [\cdots]_{+}$. In this case, the above expression translates to
\begin{equation}\label{can_3}
\begin{split}
\left[\psi_{M\alpha}(\mathbf{x},t),\dot{\psi}_{M\beta}(\mathbf{x}',t)\right]_{+} &= i\mathcal{C}^{-1}_{\alpha\beta}\delta^{(3)}(\mathbf{x}-\mathbf{x}'),
\end{split}
\end{equation}
and
\begin{equation}
\begin{split}
\left[a^{j s}_{\mathbf{p}},a^{k r\dagger}_{\mathbf{p}'}\right]_{+}&=(2\pi)^3\delta^{j k}\delta^{r s}\delta^{(3)}(\mathbf{p}-\mathbf{p}').
\end{split}
\label{canonicalrelations_2}
\end{equation}
This theory is also microcausal, satisfying
\begin{equation}\label{micro}
 \begin{split}
\left[\psi_{M\alpha}(x),\widehat{\psi}_{M\beta}(y)\right]_{+} &=\delta_{\alpha\beta}\Delta(x-y),
    \end{split}
\end{equation}
and its normal-ordered four-momentum operator  is Hermitian and bounded from below 
\begin{equation}
\begin{split}
    P^{\mu}&= :\frac{1}{2}\int  \frac{d^3\mathbf{p}}{(2\pi)^{3}}p^{\mu}\sum_{j,s}\left(a^{j s\dagger}_{\mathbf{p}}a^{j s}_{\mathbf{p}}- a^{j s}_{\mathbf{p}}a^{j s\dagger}_{\mathbf{p}}\right):=\int  \frac{d^3\mathbf{p}}{(2\pi)^{3}}p^{\mu}\sum_{j,s} a^{j s\dagger}_{\mathbf{p}}a^{j s}_{\mathbf{p}}.
\end{split}
\end{equation}
The Feynman propagator in this case is given by
\begin{equation}
\begin{split}
&S^F_{\alpha\beta}(x-y)=\bra{0}T\left[\psi_\alpha(x)\psi_\beta(y)\right]\ket{0}=\int\frac{d^4p}{(2\pi)^4}\left[\frac{i\mathcal{C}^{-1}_{\alpha\beta}}{p^2-m^2+i\epsilon}\right]e^{-i p\cdot(x-y)}.
\end{split}
\end{equation}
For a theory where the self conjugate field does not interact with other non self-conjugate fermion fields, the factors $\mathcal{C}^{-1}$ coming from the propagator and the factors $\mathcal{C}$ coming from the vertices cancel with each other, and only the factors involving $\mathcal{C}$ from external legs need to be taken into account. In this way, the Feynman rules for $\psi_M$ can be written without explicit factors of $\mathcal{C}$ and thus become equivalent to those of $\psi$.

\subsection{Symmetries}

As evident from the Lagrangian in Eq.~\eqref{Lag1}, the theory exhibits a global $U(1)$ symmetry under the phase transformations $\psi\to e^{i\theta}\psi$ and $\widehat{\psi}\to e^{-i\theta}\widehat{\psi}$ for a constant real parameter $\theta$. The associated conserved charge takes the form
\begin{equation}\label{charge}
\begin{split}
    Q &= :i\int d^3\mathbf{x}\bigg(\widehat{\psi}\dot{\psi}-\dot{\widehat{\psi}}\psi\bigg): \quad = \ :\int \frac{d^3\mathbf{p}}{(2\pi)^{3}}\sum_{j,s}\left(a^{j s\dagger}_{\mathbf{p}}a^{j s}_{\mathbf{p}}+\varsigma b^{j s}_{\mathbf{p}}b^{j s\dagger}_{\mathbf{p}}\right):\\
    &= \int \frac{d^3\mathbf{p}}{(2\pi)^{3}}\sum_{j,s}\left(a^{j s\dagger}_{\mathbf{p}}a^{j s}_{\mathbf{p}}-b^{j s\dagger}_{\mathbf{p}}b^{j s}_{\mathbf{p}}\right),
\end{split}
\end{equation}
which is explicitly Hermitian as the normal-ordered Hamiltonian.

Furthermore, the free theory hosts an extended global symmetry rooted in the equal-time graded bracket of the field and the dual field $[\psi_\alpha(x), \widehat{\psi}_\beta(x)]_{\varsigma} = 0$ (obtained from Eq.~\eqref{micro} at $x=y$). By defining the composite multiplet
\begin{equation}
    \Psi(x) = \begin{pmatrix}
    \widehat{\psi}^{\, T}(x) \\
    \psi(x)
    \end{pmatrix},
\end{equation}
the Lagrangian in Eq.~\eqref{Lag1} can be rewritten as
\begin{equation}
    \mathcal{L} = \frac{1}{2} \partial^\mu \Psi^T \Phi\,  \partial_\mu \Psi - \frac{m^2}{2} \Psi^T \Phi \, \Psi.\label{Lag2}
\end{equation}
Here, $\Phi$ is the  $8 \times 8$ matrix  written in $4 \times 4$ block form as
\begin{equation}
    \Phi = \begin{pmatrix}
    0_{4 \times 4} & 1_{4 \times 4} \\
   \shortminus \varsigma 1_{4 \times 4} & 0_{4 \times 4}
    \end{pmatrix} .
\end{equation}
Consequently, the action in Eq.~\eqref{Lag2} remains invariant under global linear mappings $\Psi \to \Psi' = T \Psi$, provided that the transformation preserves $\Phi$ according to $T^T\Phi T=\Phi$. For the fermionic case, the matrix $\Phi$ becomes the $8\times 8$ symplectic matrix and the mathematical structure is isomorphic to the symplectic group $Sp(8,\mathbb{C})$, while for the bosonic case, the resulting structure is isomorphic to the group $O(8,\mathbb{C})$.  In contrast, the self-conjugate theory of Eq.~\eqref{self_conjugate} displays a symmetry group isomorphic to
$Sp(4,\mathbb{C})$ satisfying $\psi'_M=T_M\psi_M$, with $T^T_M\mathcal{C}T_M=\mathcal{C}$. A suitable basis for its 10 generators is given by
\begin{equation}
1\otimes\boldsymbol{\sigma}, \qquad
\sigma^1\otimes 1,\qquad\sigma^2\otimes\boldsymbol{\sigma},\qquad\sigma^3\otimes\boldsymbol{\sigma}.  
\end{equation}
Finally, the behavior of the field $\psi$ and its dual under the discrete symmetries of parity ($P$), charge conjugation ($C$), and time reversal ($T$) is inherited from the transformation properties of the momentum-space operators:
\begin{gather}
    \mathrm{P}a^{js\dagger}_{\mathbf{p}}\mathrm{P}^{-1} = -i(-1)^{j-1} a^{js\dagger}_{-\mathbf{p}},\quad \mathrm{P}b^{js\dagger}_{\mathbf{p}}\mathrm{P}^{-1} = -i(-1)^{j-1}b^{js\dagger}_{-\mathbf{p}},\nonumber\\
    \mathrm{C}a^{js\dagger}_{\mathbf{p}}\mathrm{C}^{-1} = \varsigma b^{js\dagger}_{\mathbf{p}},\quad\quad\quad \mathrm{C}b^{js\dagger}_{\mathbf{p}}\mathrm{C}^{-1} = a^{js\dagger}_{\mathbf{p}},\\
    \mathrm{T}a^{js\dagger}_{\mathbf{p}}\mathrm{T}^{-1} = 2s a^{j(-s)\dagger}_{-\mathbf{p}},\quad\quad\quad \mathrm{T}b^{js\dagger}_{\mathbf{p}}\mathrm{T}^{-1} =  2s b^{j(-s)\dagger}_{-\mathbf{p}}.\nonumber
\end{gather}
These operator relations map to coordinate-space transformations on the fields as follows
\begin{gather}
    \mathrm{P}\psi(x)\mathrm{P}^{-1} = i\gamma^0\psi(\mathcal{P}x),\quad \mathrm{P}\widehat{\psi}(x)\mathrm{P}^{-1} = -i\widehat{\psi}(\mathcal{P}x)\gamma^0,\nonumber\\
    \mathrm{C}\psi(x)\mathrm{C}^{-1} =  \mathcal{C} \widehat{\psi}^{\,T},\qquad \mathrm{C}\widehat{\psi}\mathrm{C}^{-1} = \varsigma\psi^T\mathcal{C},\\
    \mathrm{T}\psi(x)\mathrm{T}^{-1} = \mathcal{C}\gamma^5 \psi(\mathcal{T}x),\qquad \mathrm{T}\widehat{\psi}\mathrm{T}^{-1} = -\widehat{\psi}(\mathcal{T}x) \gamma^5\mathcal{C},\nonumber
\end{gather}
with $\mathcal{P} = \mathrm{diag}(1,\shortminus 1,\shortminus1,\shortminus1)$, $\mathcal{T} = \mathrm{diag}(\shortminus1,1,1,1)$, and $\mathcal{C} = -i \gamma^2\gamma^0$ in the chiral representation. The pseudo-Hermitian bilinears thus behave under parity inversion as
\begin{equation}
\begin{split}
    \mathrm{P}\widehat{\psi}\psi\mathrm{P}^{-1} &= \widehat{\psi}\psi,\nonumber\\
    \mathrm{P}\widehat{\psi}i\gamma^5\psi\mathrm{P}^{-1} &= -\widehat{\psi}i\gamma^5\psi,\nonumber\\
    \mathrm{P}\widehat{\psi}\gamma^\mu\psi\mathrm{P}^{-1} &= \mathcal{P}^\mu{}_\nu\widehat{\psi}\gamma^\nu\psi,\nonumber\\
    \mathrm{P}\widehat{\psi}\gamma^\mu\gamma^5\psi\mathrm{P}^{-1} &= -\mathcal{P}^\mu{}_\nu\widehat{\psi}\gamma^\nu\gamma^5\psi,\nonumber\\
    \mathrm{P}\widehat{\psi}\sigma^{\mu\nu}\psi\mathrm{P}^{-1} &= \mathcal{P}^\mu{}_\lambda\mathcal{P}^\nu{}_\rho\widehat{\psi}\sigma^{\lambda\rho}\psi,
\end{split}
\end{equation}
with $\sigma^{\mu\nu} \equiv 2M^{\mu\nu} = i[\gamma^\mu,\gamma^\nu]/2$. Similarly, their transformation properties under charge conjugation and time reversal take the form
\begin{equation}
\begin{split}
    \mathrm{C}\widehat{\psi}\psi\mathrm{C}^{-1} &= \widehat{\psi}\psi,\nonumber\\
    \mathrm{C}\widehat{\psi}i\gamma^5\psi\mathrm{C}^{-1} &= \widehat{\psi}i\gamma^5\psi,\nonumber\\
    \mathrm{C}\widehat{\psi}\gamma^\mu\psi\mathrm{C}^{-1} &= -\widehat{\psi}\gamma^\mu\psi,\nonumber\\
    \mathrm{C}\widehat{\psi}\gamma^\mu\gamma^5\psi\mathrm{C}^{-1} &= +\widehat{\psi}\gamma^\mu\gamma^5\psi,\nonumber\\
    \mathrm{C}\widehat{\psi}\sigma^{\mu\nu}\psi\mathrm{C}^{-1} &= -\widehat{\psi}\sigma^{\lambda\rho}\psi,
\end{split}
\end{equation}
and
\begin{equation}
\begin{split}
    \mathrm{T}\widehat{\psi}\psi\mathrm{T}^{-1} &= \widehat{\psi}\psi,\nonumber\\
    \mathrm{T}\widehat{\psi}i\gamma^5\psi\mathrm{T}^{-1} &= -\widehat{\psi}i\gamma^5\psi,\nonumber\\
    \mathrm{T}\widehat{\psi}\gamma^\mu\psi\mathrm{T}^{-1} &= -\mathcal{T}^\mu{}_\nu\widehat{\psi}\gamma^\nu\psi,\nonumber\\
    \mathrm{T}\widehat{\psi}\gamma^\mu\gamma^5\psi\mathrm{T}^{-1} &= -\mathcal{T}^\mu{}_\nu\widehat{\psi}\gamma^\nu\gamma^5\psi,\nonumber\\
    \mathrm{T}\widehat{\psi}\sigma^{\mu\nu}\psi\mathrm{T}^{-1} &= -\mathcal{T}^\mu{}_\lambda\mathcal{T}^\nu{}_\rho\widehat{\psi}\sigma^{\lambda\rho}\psi.
\end{split}
\end{equation}
Together, these mappings confirm that the free theory fully preserves each discrete symmetry independently, ensuring strict $CPT$ invariance. For the self-conjugate field, all the above relations hold for Fermi--Dirac statistics by performing the replacements $\psi\rightarrow \psi_M$ and $b^{js}_{\mathbf{p}}\rightarrow a^{js}_{\mathbf{p}}$. This implies that the vector and tensor bilinears built from $\psi_M$ vanish identically
\begin{equation}
\widehat{\psi}_M\gamma^\mu\psi_M=0,\qquad \widehat{\psi}_M\sigma^{\mu\nu}\psi_M=0.
\end{equation}


\section{Self-Interactions}\label{sec:Fierz}
Since our fields are of mass dimension one, our SOPHIES admit renormalizable quartic self-interactions. In this section, we identify the set  of independent self-interactions available for each one. The quartic self-interactions for our spin-$1/2$ fields are related through Fierz identities as
\begin{equation}
\begin{aligned}
\varsigma\left(
\widehat{\psi}
\psi
\right)^2
=&
-
\frac{1}{4}
\left(\widehat{\psi}\psi\right)^2
-
\frac{1}{4}
\left(
\widehat{\psi}\gamma^{5}\psi
\right)^2
-
\frac{1}{4}
\left(
\widehat{\psi}\gamma^{\mu}\psi
\right)^2
+
\frac{1}{4}
\left(
\widehat{\psi}\gamma^{5}\gamma^{\mu}\psi
\right)^2
-
\frac{1}{8}
\left(
\widehat{\psi}\sigma^{\mu\nu}\psi
\right)^2,
\\
\varsigma
\left(
\widehat{\psi}
\gamma^{5}
\psi
\right)^2
=&
-
\frac{1}{4}
\left(\widehat{\psi}\psi\right)^2
-
\frac{1}{4}
\left(
\widehat{\psi}\gamma^{5}\psi
\right)^2
+
\frac{1}{4}
\left(
\widehat{\psi}
\gamma^{\mu}
\psi
\right)^2
-
\frac{1}{4}
\left(
\widehat{\psi}
\gamma^{5}
\gamma^{\mu}
\psi
\right)^2
-
\frac{1}{8}
\left(
\widehat{\psi}
\sigma^{\mu\nu}
\psi
\right)^2,
\\
\varsigma\left(
\widehat{\psi}
\gamma^{\mu}
\psi
\right)^2
=&
-
\left(\widehat{\psi}\psi\right)^2
+
\left(
\widehat{\psi}
\gamma^{5}
\psi
\right)^2
+
\frac{1}{2}
\left(
\widehat{\psi}
\gamma^{\mu}
\psi
\right)^2
+
\frac{1}{2}
\left(
\widehat{\psi}
\gamma^{5}
\gamma^{\mu}
\psi
\right)^2,
\\
\varsigma\left(
\widehat{\psi}
\gamma^{5}
\gamma^{\mu}
\psi
\right)^2
=&
+
\left(\widehat{\psi}\psi\right)^2
-
\left(
\widehat{\psi}\gamma^{5}\psi
\right)^2
+
\frac{1}{2}
\left(
\widehat{\psi}
\gamma^{\mu}
\psi
\right)^2
+
\frac{1}{2}
\left(
\widehat{\psi}\gamma^{5}\gamma^{\mu}\psi
\right)
^2,
\\
\varsigma\left(
\widehat{\psi}
\sigma^{\mu\nu}
\psi
\right)^2
=&
-
3
\left(\widehat{\psi}\psi\right)^2
-
3
\left(
\widehat{\psi}\gamma^{5}\psi
\right)^2
+
\frac{1}{2}
\left(
\widehat{\psi}\sigma^{\mu\nu}\psi
\right)^2,\label{eq:fierz_1}
\end{aligned}
\end{equation}
with $\varsigma=+1$ for F-SOPHY and $\varsigma=-1$ for B-SOPHY. This system can be solved for F-SOPHY in terms of the scalar, pseudo-scalar, and vector bilinears to obtain
\begin{equation}
\begin{aligned}
\left(
\widehat{\psi}
\gamma^{5}
\gamma^{\mu}
\psi
\right)^2
=&
2
\left(
\widehat{\psi}
\psi
\right)^2
-
2
\left(
\widehat{\psi}
\gamma^{5}
\psi
\right)^2
+
\left(
\widehat{\psi}
\gamma^{\mu}
\psi
\right)^2,
\\
\left(
\widehat{\psi}
\sigma^{\mu\nu}
\psi
\right)^2
=&
-6
\left(
\widehat{\psi}
\psi
\right)^2
-
6
\left(
\widehat{\psi}
\gamma^{5}
\psi
\right)^2.
\end{aligned}
\end{equation}
Thus, for F-SOPHY, there are only three independent self-interactions:
\begin{equation}
\mathcal{L}_{{F}}^{\text{self}}=\frac{\lambda_1}{2}\left(
\widehat{\psi}
\psi
\right)^2+\frac{\lambda_2}{2}\left(
\widehat{\psi}
\gamma^{5}
\psi
\right)^2+\frac{\lambda_3}{2}\left(
\widehat{\psi}
\gamma^{\mu}
\psi
\right)^2.\label{eq:selfF}
\end{equation}
Similarly, for B-SOPHY, Eqs.~\eqref{eq:fierz_1} yield
\begin{equation}
\begin{aligned}
\left(
\widehat{\psi}
\gamma^{\mu}
\psi
\right)^2
=&
\left(
\widehat{\psi}
\psi
\right)^2
-
\left(
\widehat{\psi}
\gamma^{5}
\psi
\right)^2,
\\
\left(
\widehat{\psi}
\gamma^{5}
\gamma^{\mu}
\psi
\right)^2
=&
-
\left(
\widehat{\psi}
\psi
\right)^2
+
\left(
\widehat{\psi}
\gamma^{5}
\psi
\right)^2,
\\
\left(
\widehat{\psi}
\sigma^{\mu\nu}
\psi
\right)^2
=&
\ 2
\left(
\widehat{\psi}
\psi
\right)^2
+
2
\left(
\widehat{\psi}
\gamma^{5}
\psi
\right)^2,
\end{aligned}
\end{equation}
and therefore, in this case there are only two independent self-interactions:
\begin{equation}
\mathcal{L}_{{B}}^{\text{self}}=\frac{\lambda_1}{2}\left(
\widehat{\psi}
\psi
\right)^2+\frac{\lambda_2}{2}\left(
\widehat{\psi}
\gamma^{5}
\psi
\right)^2.\label{eq:selfB}
\end{equation}

For M-SOPHY, the vector, and tensor bilinears vanish identically, and Eqs.~\eqref{eq:fierz_1} reduce to
\begin{equation}
\begin{aligned}
\left(
\widehat{\psi}_M
\psi_M
\right)^2
=&
-
\frac{1}{4}
\left(\widehat{\psi}_M\psi_M\right)^2
-
\frac{1}{4}
\left(
\widehat{\psi}_M\gamma^{5}\psi_M
\right)^2
+
\frac{1}{4}
\left(
\widehat{\psi}_M\gamma^{5}\gamma^{\mu}\psi_M
\right)^2
,
\\
\left(
\widehat{\psi}_M
\gamma^{5}
\psi_M
\right)^2
=&
-
\frac{1}{4}
\left(\widehat{\psi}_M\psi_M\right)^2
-
\frac{1}{4}
\left(
\widehat{\psi}_M\gamma^{5}\psi_M
\right)^2
-
\frac{1}{4}
\left(
\widehat{\psi}_M
\gamma^{5}
\gamma^{\mu}
\psi_M
\right)^2,
\\
\left(
\widehat{\psi}_M
\gamma^{5}
\gamma^{\mu}
\psi_M
\right)^2
=&
+
\left(\widehat{\psi}_M\psi_M\right)^2
-
\left(
\widehat{\psi}_M\gamma^{5}\psi_M
\right)^2
+
\frac{1}{2}
\left(
\widehat{\psi}_M\gamma^{5}\gamma^{\mu}\psi_M
\right)
^2.
\label{eq:fierz_M}
\end{aligned}
\end{equation}
This system can be solved in terms of the scalar bilinear only, according to
\begin{equation}
\begin{aligned}
\left(
\widehat{\psi}_M\gamma^{5}\psi_M
\right)^2=&-\left(
\widehat{\psi}_M\psi_M
\right)^2,\\
\left(
\widehat{\psi}_M\gamma^{5}\gamma^\mu\psi_M
\right)^2=&\, 4\left(
\widehat{\psi}_M\psi_M
\right)^2,
\end{aligned}
\end{equation}
implying that in M-SOPHY the only independent self-interaction is
\begin{equation}
\mathcal{L}_{M}^{\text{self}}=\frac{\lambda_1}{2}\left(
\widehat{\psi}_M
\psi_M
\right)^2.
\end{equation}

\section{One-loop Renormalization }\label{sec:ren}

\subsection{Feynman rules}
In this section, we study the one-loop renormalization of the F- and B-SOPHIES  coupled to an abelian $U(1)$  gauge boson and enhanced with a Pauli term and including self-interactions.  This model was originally studied in~\cite{Vaquera-Araujo:2012jlk} in the context of the unphysical Hermitian fermionic theory analogous to Eq.~\eqref{Lag1} but with $\widehat{\psi}$ replaced by $\overline{\psi}$.  The first observation we can make is that the model presented in~\cite{Vaquera-Araujo:2012jlk} is overcomplete, since in that case only two of the three self-interactions analyzed there are independent, according to the analysis presented in the previous section. We also employ the results obtained to analyze the one-loop renormalization of the only independent quartic self-interaction of M-SOPHY.
\begin{figure}[ht]
\begin{center}
\begin{tikzpicture}[scale=1.0]
\draw[fermion] (-6,4)  -- node[above]{$p$}(-4,4) ;
\node at (-6,4.25){$b$};
\node at (-4,4.25){$a$};
\node at (0,4) {$iS_{ab}(p) =i\frac{\mathds{1}_{ab}}{p^{2}-m^{2}} $};
\draw[decorate,decoration={snake,segment length=5,amplitude=1}] (-6,2) --
node[above]{$q$}(-4,2);
\node at (-6,2.2) {$\mu $};
\node at (-4,2.2) {$\nu $};
\node at (0,2) {$ i\Delta^{\mu\nu}(q) =
i\frac{-g^{\mu \nu }+\left(1-\xi\right)\frac{q^\mu q^\nu}{q^2}}{q^{2}+i\varepsilon } $};
\draw[fermion] (-6,-2+.5) --(-5,-1+.5) ;
\draw[fermion] (-5,-1+.5)  --  (-4,-2+.5) ;
\draw[decorate,decoration={snake,segment length=5,amplitude=1}]
(-5,0+.5)--(-5,-1+.5);
\node at (-5,0.75) {$q, \mu $};
\node at (-6,-1.75){$p,b$};
\node at (-4,-1.75) {$p^{\prime},a$};
\node at (0,-0.75+.5) {$-ieV^{\mu }_{ab}(p,p^{\prime}) =$};
\node at (0,-1.25+.5) {$-ie
\left[(p-p^{\prime})^{\mu }\mathds{1}_{ab}-i\frac{g}{2} \sigma^{\mu \nu }_{ab}(p+p^{\prime})_{\nu}\right] $};
\draw[decorate,decoration={snake,segment length=5,amplitude=1}]
(-6,-3)--(-5,-4);
\draw[decorate,decoration={snake,segment length=5,amplitude=1}]
(-5,-4)--(-4,-3);
\draw[fermion] (-6,-5)  --(-5,-4) ;
\draw[fermion] (-5,-4)  --  (-4,-5);
\node at (-6,-5.25){$p,b$};
\node at (-4,-5.25) {$p^{\prime},a$};
\node at (-6,-2.75) {$q,\mu $};
\node at (-4,-2.75) {$q',\nu $};
\node at (0,-4) { $\,\,\,\,\,\,\, ie^2V^{\mu \nu}_{ab}=2ie^{2} g^{\mu \nu }\mathds{1}_{ab} $};
\draw[fermion] (-6,-7+.5)--(-5,-7.5+.5);
\draw[fermion] (-5,-7.5+.5)--(-4,-7+.5);
\draw[fermion] (-6,-9+.5)  --(-5,-8.5+.5) ;
\draw[fermion] (-5,-8.5+.5)  --  (-4,-9+.5);
\draw[scalar] (-5,-7.5+.5) --(-5,-8.5+.5);
\node at (-6,-9.25+.5){$p_{2},d$};
\node at (-4,-9.25+.5) {$p^{\prime}_{2},c$};
\node at (-6,-6.75+.5) {$p_{1},b$};
\node at (-4,-6.75+.5) {$p^{\prime}_{1},a$};
\node at (0,-7.5+.25) { $i\lambda_{abcd}=$};
\node at (0,-8+.25) { $i[\lambda_{1}
\mathds{1}_{ab}\mathds{1}_{cd}+\lambda_{2}\gamma^5_{ab}\gamma^5_{cd}+\vartheta\lambda_{3} \gamma^{\mu}_{ab}\gamma_{\mu\,cd}]$};
\end{tikzpicture}
\end{center}
\caption{Feynman rules for the non-minimally pseudo-Hermitian $U(1)$ gauge theory in $R_\xi$-gauge, enhanced with a mass-dimension-four Pauli term and self-interactions. Spinor indices run from 1 to 4 and Lorentz indices from 0 to 3. Here, $\mathds{1}_{ab}$ is the
Kronecker delta symbol, and all momenta are incoming.}
\label{fig:FD}
\end{figure}

The non-minimally coupled $U(1)$ gauge invariant pseudo-Hermitian Lagrangian for fields transforming in the $(1/2,0)\oplus(0,1/2)$ representation of the RLG is
\begin{equation}
\begin{split}
\mathscr{L}=&  -\frac{1}{4}F^{\mu\nu}F_{\mu\nu}-\frac{\left(\partial^\mu A_\mu\right)^2}{2\xi}+ D^{\dagger\mu} \widehat{\psi}
\left(g_{\mu\nu}- i\frac{g}{2} \sigma_{\mu\nu} \right)D^\nu\psi-m^2\widehat{\psi}\psi +\mathcal{L}^{\text{self}},
\label{U(1)lag}
\end{split}
\end{equation}
with covariant derivative $D_\mu= \partial_\mu+ie A_\mu $, $e$ as the $U(1)$ coupling constant, $g$ as the Pauli coupling, and $\mathcal{L}^{\text{self}}$ as the self-interaction Lagrangian adequate for fermion or boson fields in Eqs.~\eqref{eq:selfF} and~\eqref{eq:selfB}, namely
\begin{equation}
\begin{split}
\mathcal{L}^{\text{self}}&=\frac{\lambda_1}{2}\left(
\widehat{\psi}
\psi
\right)^2+\frac{\lambda_2}{2}\left(
\widehat{\psi}
\gamma^{5}
\psi
\right)^2+\vartheta\frac{\lambda_3}{2}\left(
\widehat{\psi}
\gamma^{\mu}
\psi
\right)^2.
\end{split}
\end{equation}
Here $\vartheta=\Theta(\varsigma)$, where $\Theta$ is the Heaviside step function, with $\varsigma=+1$ for F-SOPHY and $\varsigma=-1$ for B-SOPHY. 
The corresponding Feynman rules for both SOPHIES are shown in Fig.~\ref{fig:FD}.
For closed fermion loops, the familiar factor of $-1$ given by Fermi statistics must be included, as well as a relative $-1$ factor between diagrams differing by the exchange of a pair of fermion lines.

In this work, we use dimensional regularization with $d=4-2\epsilon$ and the naive prescription for the
chirality operator $\gamma^5$, which assumes
\begin{equation}
[\gamma^5,\sigma^{\mu\nu}]=0,\qquad\{\gamma^5,\gamma^{\mu}\}=0, 
\end{equation}
in $d$ dimensions.
 Although this approach is not self-consistent, it does not lead to inconsistencies since $\gamma^5$ appears in pairs for all the processes
involved and can be safely used~\cite{Chanowitz:1979zu}.  

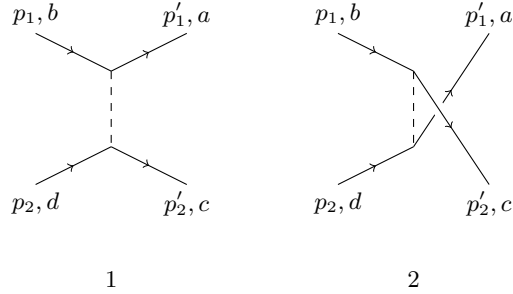
\begin{figure}
    \centering
\begin{tikzpicture}[scale=1.0]
\draw[fermion] (-6,-7+.5)--(-5,-7.5+.5);
\draw[fermion] (-5,-7.5+.5)--(-4,-7+.5);
\draw[fermion] (-6,-9+.5)  --(-5,-8.5+.5) ;
\draw[fermion] (-5,-8.5+.5)  --  (-4,-9+.5);
\draw[scalar] (-5,-7.5+.5) --(-5,-8.5+.5);
\node at (-6,-9.25+.5){$p_{2},d$};
\node at (-4,-9.25+.5) {$p^{\prime}_{2},c$};
\node at (-6,-6.75+.5) {$p_{1},b$};
\node at (-4,-6.75+.5) {$p^{\prime}_{1},a$};
\node at (-5,-10+.25) {$1$};
\draw[fermion] (-6+4,-7+.5)--(-5+4,-7.5+.5);
\draw[fermion] (-6+4,-9+.5)  --(-5+4,-8.5+.5) ;
\draw[fermion] (-5+4,-8.5+.5)  --  (-4+4,-7+.5);
\draw[draw=white,line width=5] (-5+4,-7.5+.5)--(-4+4,-9+.5);
\draw[fermion] (-5+4,-7.5+.5)--(-4+4,-9+.5);
\draw[scalar] (-5+4,-7.5+.5) --(-5+4,-8.5+.5);
\node at (-6+4,-9.25+.5){$p_{2},d$};
\node at (-4+4,-9.25+.5) {$p^{\prime}_{2},c$};
\node at (-6+4,-6.75+.5) {$p_{1},b$};
\node at (-4+4,-6.75+.5) {$p^{\prime}_{1},a$};
\node at (-5+4,-10+.25) {$2$};
\end{tikzpicture}
    \caption{ Feynman diagrams contributing to the tree-level four-spinor vertex function for F- and B-SOPHIES.}
    \label{fig:selfint}
\end{figure}

In our representation of Feynman diagrams for self-interactions, we adopt the prescription of~\cite{Bondi:1989nq},  where the four-spinor vertex is drawn as a  dashed line joining two spinor lines, as  depicted in Fig.~\ref{fig:FD}. This allows for automatic accounting of the different possible contractions
between spin-$1/2$ fields. With this convention, the self-interaction contribution to the tree-level
four-spinor function is given by 
the diagrams of Fig.~\ref{fig:selfint}, and yields $i(\lambda_{abcd}-\lambda_{cbad})$ for fermion lines and $i(\lambda_{abcd}+\lambda_{cbad})$ for boson lines.\\

\begin{figure}[ht]
\begin{center}
\begin{tikzpicture}[scale=1.0]
\draw[fermion] (-6,4)  --(-5,4) ;
\draw[fermion] (-5,4)  --(-4,4) ;
\node at (-5,4) {$\bullet $};
\node at (-6,4.25){$b$};
\node at (-5,4.25){$p$};
\node at (-4,4.25){$a$};
\node at (0,4) {$ i(p^2-m^2) \delta_{2}\mathds{1}_{ab} - im^2 \delta_m\mathds{1}_{ab}$};
\draw[decorate,decoration={snake,segment length=5,amplitude=1}] (-6,2) --
node[above]{$q$}(-4,2);
\node at (-5,2) {$\bullet $};
\node at (-6,2.2) {$\mu $};
\node at (-4,2.2) {$\nu $};
\node at (0,2) {$-i(g^{\mu\nu} q^2 -q^\mu q^\nu )\delta_{1}   $};
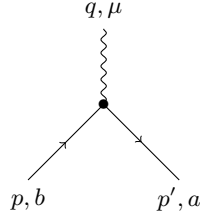
\draw[fermion] (-6,-2+.5) --(-5,-1+.5) ;
\draw[fermion] (-5,-1+.5)  --  (-4,-2+.5) ;
\draw[decorate,decoration={snake,segment length=5,amplitude=1}]
(-5,0+.5)--(-5,-1+.5);
\node at (-5,-.5) {$\bullet $};
\node at (-5,0.75) {$q, \mu $};
\node at (-6,-1.75){$p,b$};
\node at (-4,-1.75) {$p^{\prime},a$};
\node at (0,-0.5) {$-ie \left[V_{ab}^{\mu}(p,p')
\delta_{e}  -i\frac{g}{2}\sigma^{\mu\nu}_{ab} (p'+p)_\nu \delta_g  \right]$};
\draw[decorate,decoration={snake,segment length=5,amplitude=1}]
(-6,-3)--(-5,-4);
\draw[decorate,decoration={snake,segment length=5,amplitude=1}]
(-5,-4)--(-4,-3);
\draw[fermion] (-6,-5)  --(-5,-4) ;
\draw[fermion] (-5,-4)  --  (-4,-5);
\node at (-5,-4) {$\bullet $};
\node at (-6,-5.25){$p,b$};
\node at (-4,-5.25) {$p^{\prime},a$};
\node at (-6,-2.75) {$q,\mu $};
\node at (-4,-2.75) {$q^{\prime},\nu $};
\node at (0,-4) { $ie^{2} V^{\mu \nu }_{ab} \delta_3 $};
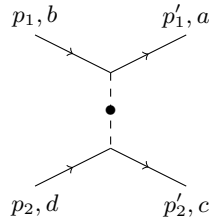
\draw[fermion] (-6,-7+.5)--(-5,-7.5+.5);
\draw[fermion] (-5,-7.5+.5)--(-4,-7+.5);
\draw[fermion] (-6,-9+.5)  --(-5,-8.5+.5) ;
\draw[fermion] (-5,-8.5+.5)  --  (-4,-9+.5);
\draw[scalar] (-5,-7.5+.5) --(-5,-8.5+.5);
\node at (-5,-7.5) {$\bullet $};
\node at (-6,-9.25+.5){$p_{2},d$};
\node at (-4,-9.25+.5) {$p^{\prime}_{2},c$};
\node at (-6,-6.75+.5) {$p_{1},b$};
\node at (-4,-6.75+.5) {$p^{\prime}_{1},a$};
\node at (0,-7.75+.25) { $i[\lambda_{1}\delta_{\lambda_1}
\mathds{1}_{ab}\mathds{1}_{cd}+\lambda_{2}\delta_{\lambda_2}\gamma^5_{ab}\gamma^5_{cd}+\vartheta\lambda_{3}\delta_{\lambda_3} \gamma^{\mu}_{ab}\gamma_{\mu cd}]$};
\end{tikzpicture}
\end{center}
\caption{Counterterm Feynman rules for the spin-$1/2$ SOPHIES.}
\label{fig:CT}
\end{figure}

The bare Lagrangian is parameterized by the field mass $m_0$, the $U(1)$ coupling $e_0$, the Pauli coupling $g_0$ and the self-interaction couplings $\lambda_{0j}$, $j=1,2,3$.  The renormalized
fields are related to the bare ones, as
 \begin{equation}
 A^{\mu} =  Z_1^{-\frac{1}{2}} A^{\mu}_0, \qquad  \psi =  Z_2^{-\frac{1}{2}} \psi_0 .
 \end{equation}
The Lagrangian can be split into its free and interacting parts
 \begin{equation}
 \mathcal{L}=\mathcal{L}_0+\mathcal{L}_{\text{int}},
 \end{equation}
with
\begin{eqnarray}
\mathcal{L}_0&=&-\frac{1}{4}F^{\mu\nu}_{0}F_{0\mu\nu}- \frac{1}{2\xi_0}(\partial^\mu A_{0\mu})^2
+ \partial^\mu \widehat{\psi}_0 \partial_{\mu} \psi_0 -m_{0}^{2} \widehat{\psi}_0 \psi_0,\\
\mathcal{L}_{\text{int}}&=& -ie_0 \left[ \widehat{\psi}_0   \left(g^{\mu\nu}+ i\frac{g_0}{2} \sigma^{\mu\nu} \right) \partial_\mu \psi_0 - \partial_\mu
\widehat{\psi}_0 \left(g^{\mu\nu}- i\frac{g_0}{2} \sigma^{\mu\nu} \right)\psi_0   \right] A_{0\nu} + e^{2}_0 \widehat{\psi}_0  \psi_0 A^\mu_0 A_{0\mu} \nonumber\\
&&+ \frac{\lambda_{01}}{2}\left(\widehat{\psi}_0\psi_0\right)^{2}+
\frac{\lambda_{02}}{2}\left(\widehat{\psi}_0\gamma^5\psi_0\right)^{2}+\vartheta
\frac{\lambda_{03}}{2}\left(\widehat{\psi}_0 \gamma^{\mu}\psi_0\right)^{2} .\nonumber
\end{eqnarray}
In terms of the renormalized fields, the free Lagrangian becomes
\begin{align}
\mathcal{L}_0=&-\frac{1}{4}F^{\mu\nu} F_{\mu\nu}  - \frac{1}{2\xi}(\partial^\mu A_{\mu})^2
-\frac{1}{4}F^{\mu\nu} F_{\mu\nu} \delta_{1} \\%
&+ \partial^\mu \widehat{\psi} \partial_{\mu} \psi   - m^{2} \widehat{\psi} \psi
+[ \partial^\mu \widehat{\psi} \partial_{\mu} \psi   - m^2   \widehat{\psi} \psi ] \delta_{2}
- \delta_{m} m^2  \hat{\psi} \psi,\nonumber
\end{align}
where we have defined
\begin{equation}\label{count1}
\delta_{1}\equiv Z_{1}-1, \qquad \delta_{2}\equiv Z_{2}-1,  \qquad \delta _{m}\equiv Z_{m}-Z_{2},\qquad Z_{m}\equiv\frac{m^2_0}{m^2} Z_{2},\qquad \xi= Z_1^{-1}\xi_0.
\end{equation}
Analogously, the interacting Lagrangian can be cast into
\begin{eqnarray}
\mathcal{L}_{\text{int}}  &=& -ie \left[ \widehat{\psi}   \left(g^{\mu\nu}+ i\frac{g}{2} \sigma^{\mu\nu} \right) \partial_\mu \psi - \partial_\mu
\widehat{\psi} \left(g^{\mu\nu}- i\frac{g}{2} \sigma^{\mu\nu} \right)\psi   \right] A_{\nu} + e^{2} \widehat{\psi}  \psi A^\mu A_{\mu} \nonumber\\
&&+ \frac{\lambda_{1}}{2}\left(\widehat{\psi}\psi\right)^{2}+
\frac{\lambda_{2}}{2}\left(\widehat{\psi}\gamma^5\psi\right)^{2}+
\vartheta\frac{\lambda_{3}}{2}\left(\widehat{\psi} \gamma^{\mu}\psi\right)^{2}  \nonumber \\ %
&& -ie\left[ \widehat{\psi}    \left(g^{\mu\nu}+ i\frac{g}{2} \sigma^{\mu\nu} \right) \partial_\mu \psi - \partial_\mu \widehat{\psi}
\left(g^{\mu\nu}- i\frac{g}{2} \sigma^{\mu\nu} \right) \psi   \right] A_{\nu} \delta_{e}\nonumber\\&&
+ \frac{e g}{2} \left[ \widehat{\psi}   \sigma^{\mu\nu}  \partial_\mu \psi+\partial_\mu \widehat{\psi} \sigma^{\mu\nu}\psi   \right] A_{\nu}   \delta_{g}  +  e^{2} \widehat{\psi}  \psi A^\mu A_{\mu} \delta_3 \nonumber\\
&&+ \frac{\delta_{\lambda_1}\lambda_{1}}{2}\left(\widehat{\psi} \psi\right)^{2}
+ \frac{\delta_{\lambda_2}\lambda_{2}}{2}\left(\widehat{\psi}\gamma^5 \psi\right)^{2}
 + \vartheta\frac{\delta_{\lambda_3}\lambda_{3}}{2}\left(\widehat{\psi}\gamma^{\mu} \psi\right)^{2},\nonumber
\end{eqnarray}
with
\begin{equation}
\delta_e\equiv Z_e - 1,  \quad
\delta_3\equiv Z_3-1, \quad
\delta_{\lambda_j}\equiv  Z_{\lambda_j}-1,\quad \delta_g\equiv Z_{eg}-Z_e ,
\label{ctc}
\end{equation}
and
\begin{equation}
Z_e\equiv\frac{e_0}{e} Z_1^{\frac{1}{2}}Z_2 ,  \quad
Z_3\equiv  \frac{e_0^2}{e^2} Z_1 Z_2, \quad
Z_{\lambda_j}\equiv \frac{\lambda_{0j}}{\lambda_{j}}Z_{2}^{2},\quad Z_{eg}\equiv \frac{g_0}{g} Z_e .
\label{ctc2}
\end{equation}
In  $d=4-2\epsilon$ dimensions, the renormalized parameters must be
modified as follows:
\begin{equation}
e \to \mu^\epsilon e,
\qquad
\lambda_{j} \to \mu^{2\epsilon}\lambda_{j},
\qquad
g \to g,
\qquad
m \to m,
\end{equation}
where $\mu$ denotes the renormalization scale. The Feynman rules for renormalized fields are given in Fig.~\ref{fig:FD}, and the corresponding rules for the counterterms are shown in Fig.~\ref{fig:CT}. In the following subsections, we will compile the results obtained for the calculation of all the divergent processes, showing that all the divergences can be absorbed successfully into the given set of counterterms provided by the theory. We adopt the on-shell renormalization scheme for all vertex functions, except for the four-spinor vertex, where we implement the modified minimal subtraction scheme ($\overline{\text{MS}}$).  Therefore, it is worth mentioning that our conventions for the Passarino--Veltman scalar integrals are as follows
\begin{eqnarray}
&& A_{0}(m^2)=\frac{(2\pi\mu)^{4-d}}{i\pi^2}      \int d^dl
 \frac{1 }{l^{2}-m^2} ,\\
&& B_{0}(p^2,m_{1}^2,m_{2}^2 )=\frac{(2\pi\mu)^{4-d}}{i\pi^2}      \int d^dl
 \frac{1 }{[l^{2}-m_{1}^2][(l+p)^2-m_{2}^2]} ,\\
 &&C_{0}(p^2_1,p^2_2;m_{1}^2,m_{2}^2,m_{3}^2)=\nonumber\\&&\quad\frac{(2\pi\mu)^{4-d}}{i\pi^2}      \int d^dl
 \frac{1 }{[l^{2}-m_{1}^2][(l+p_1)^2-m_{2}^2][(l+p_2)^2-m_{3}^2]}.\\
\end{eqnarray}
The scalar 1-point integral is given by
\begin{equation}
A_{0}(m^2)=m^2\left[\frac{1}{\tilde{\epsilon}}-\ln\left(\frac{m^2}{\mu^2}\right)+1\right],
\end{equation}
with
\begin{equation}
 \frac{1}{\tilde{\epsilon}} \equiv \frac{1}{\epsilon} -\gamma_E+ \ln{4\pi},
\end{equation}
where $\gamma_{E}$ is the  Euler--Mascheroni constant. Similarly, using the Feynman parametrization, the scalar two-point function $B_0$ can be decomposed into its divergent and finite parts as
\begin{equation}
 B_{0}(p^2,m_{1}^2,m_{2}^2 )=  \frac{1}{\tilde{\epsilon}}+
 \tilde{B}_{0}(p^2,m_{1}^2,m_{2}^2 ),
\label{B0explicit}
\end{equation}
with
\begin{equation}\label{B0explicit1}
 \tilde{B}_{0}(p^2,m_{1}^2,m_{2}^2 )\equiv - \int^1_0 dx
  \ln \left[\frac{m_{1}^{2}(1-x)+ m_{2}^2 x   - p^2 x(1-x)}{\mu^{2}} \right].
\end{equation}

\subsection{Vacuum polarization}
At one-loop order, the vacuum polarization receives contributions from the two diagrams shown in Fig.~\ref{fig:polava}. Using the Feynman rules derived from Fig.~\ref{fig:FD} and including
the corresponding counterterm shown in Fig.~\ref{fig:CT}, the vacuum polarization is given by
\begin{equation}
-i\Pi^{\mu\nu}(q)=-i\Pi^{*\mu\nu}(q)-i\delta_{1} \left(g^{\mu\nu}q^2-q^{\mu}q^{\nu}\right),
\label{vacuo}
\end{equation}
where $-i\Pi^{*\mu\nu}(q)$ denotes the amplitude arising from the one-loop diagrams in Fig.~\ref{fig:polava}:
\begin{figure}[t]
\begin{center}
\begin{tikzpicture}
\draw[decorate,decoration={snake,segment length=5,amplitude=1}] (-5,0)node[above]{$q,\mu$}  --
(-4,0);
\draw[fermion] (-2.5,0) arc (0:180:.75);
\draw[fermion] (-4,0) arc (-180:0:0.75);
\draw[decorate,decoration={snake,segment length=5,amplitude=1}] (-2.5,0)  --(-1.5,0) node[above]{$q,\nu$};
\node at (-3.25,1.1) {$l$};
\node at (-3.25,-.4) {$l+q$};
\node at (-3.25,-1.5) {$1$};
\draw[decorate,decoration={snake,segment length=5,amplitude=1}] (.5,0)node[above]{$q,\mu$} -- (3.5,0)node[above]{$q,\nu$};
\draw[fermionloop] (2,.75) circle (.70) ;
\node at (2,.75) {$l$};
\node at (2,-1.5) {$2$};
\end{tikzpicture}
\end{center}\caption{Feynman diagrams for the vacuum polarization at one-loop level.}
\label{fig:polava}
\end{figure}
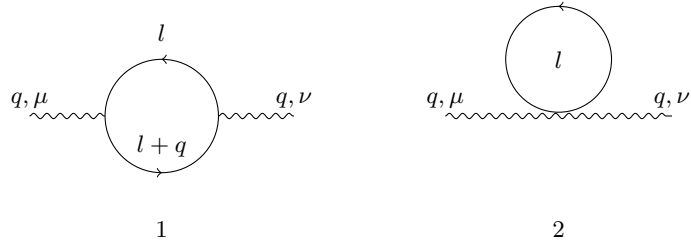

\begin{equation}
-i\Pi^{*\mu\nu}(q)=-i (g^{\mu\nu}q^2-q^\mu q^\nu) \pi^{*}(q^2),
\end{equation}
with
\begin{equation} \label{vpest}
\pi^{*}(q^2) =    \frac{  e^2\tau } {24\pi^2} \left\{\frac{3 g^2-4}{8}
 B_{0}(q^2,m^2,m^2 )  +  \frac{2 m^{2}}{q^2}\left[  B_{0}(q^2,m^2,m^2 )  - B_{0}(0,m^2,m^2 )   \right]
 -\frac{1}{3} \right\}.
\end{equation}
Here $\tau=\varsigma\Tr[\mathds{1}]=4\varsigma$, with $\varsigma=1$ for F-SOPHY and  $\varsigma=-1$ for B-SOPHY. In the following, we will leave $\tau$ unevaluated, since it will become useful to compare our results to those of other theories. Thus, the  total renormalized vacuum polarization tensor can be written as
\begin{equation}
-i\Pi^{\mu\nu}(q)= -i(g^{\mu\nu}q^2-q^\mu q^\nu) \pi(q^2) ,
\end{equation}
where $\pi(q^2)$ is the renormalized form factor,
\begin{equation}
\pi(q^2)=\pi^{*}(q^2)+\delta_{1}.
\end{equation}
In the on-shell renormalization scheme, the following condition is imposed:
\begin{equation}
\pi(q^2\to 0)=0.
\label{con1r}
\end{equation}
This  condition determines the counterterm $\delta_{1}$ as
\begin{equation}
\delta_{1}= - \pi^{*}(q^2=0)=  - \frac{  e^2\tau } {(4\pi)^2}\left\{
\left[\frac{1}{\tilde\epsilon}- \ln{\frac{m^2}{\mu^2}}\right]\left( \frac{3g^2-4}{12}\right)-\frac{2}{9}\right\},
\label{condr11}
\end{equation}
and the resulting renormalized form factor becomes
\begin{equation}
\pi(q^2)  =   \frac{  e^2 \tau } {24\pi^2} 
\left( \frac{3 g^2-4}{8} +  \frac{2 m^{2}}{q^2}\right)\left[  B_{0}(q^2,m^2,m^2 )  - B_{0}(0,m^2,m^2 )   \right] .
\end{equation}
Notice that for $g=2$ in F-SOPHY, this form factor is exactly double that of its Dirac QED counterpart, consistent with the fact that our fields have double the degrees of freedom with respect to a Dirac spinor. This behavior will be a constant during the calculations for F-SOPHY with $g=2$, $\lambda_i=0$, $i=1,2,3$.
\subsection{Spinor Self-Energy}
\begin{figure}
    \centering
\begin{tikzpicture}
\draw[fermion] (-4.5,0) node[above]{$p,b$}  --(-3.75,0);
\draw[fermion] (-3.75,0)  --(-2.25,0);
\draw[fermion] (-2.25,0)  --(-1.5,0) node[above]{$p,a$};
\draw [decorate,decoration={snake,segment length=2*3.141592*.75,amplitude=1}](-2.25,0) arc (0:180:0.75);
\node at (-3,1.05) {$l$};
\node at (-3,.25) {$l+p$};
\node at (-3,-.5) {$1$};
\draw[fermion] (-.5+1,0)node[above]{$p,b$} -- (1+1,0);
\draw[fermion] (1+1,0) -- (2.5+1,0)node[above]{$p,a$};
\draw[decorate,decoration={snake,segment length=2*3.141592*.75,amplitude=1}] (2,0) arc (-90:275:0.75);
\node at (1+1,.75) {$l$};
\node at (-3+5,-.5) {$2$};
\draw[fermion] (-4.5,-3) node[above]{$p,b$}  --(-3.75,-3);
\draw[fermion] (-3.75,-3)  --(-2.25,-3);
\draw[fermion] (-2.25,-3)  --(-1.5,-3) node[above]{$p,a$};
\draw [scalar](-2.25,-3) arc (0:180:0.75);
\node at (-3,1.05-3) {$l$};
\node at (-3,.25-3) {$l+p$};
\node at (-3,-3.5) {$3$};
\draw[fermion] (3.5-3,0-3) node[above]{$p,b$}--(5-3,0-3) ;
\draw[fermionloop] (5-3,1-3) circle (.50);
\draw[fermion] (5-3,0-3)--(6.5-3,0-3) node[above]{$p,a$} ;
\draw[scalar] (5-3,.5-3)--(5-3,0-3);
\node at (5-3,1-3) {$l$};
\node at (-3+5,-.5-3) {$4$};
\end{tikzpicture}
    \caption{Feynman diagrams for
the spinor self-energy at
one-loop level.}
    \label{fig:self}
\end{figure}
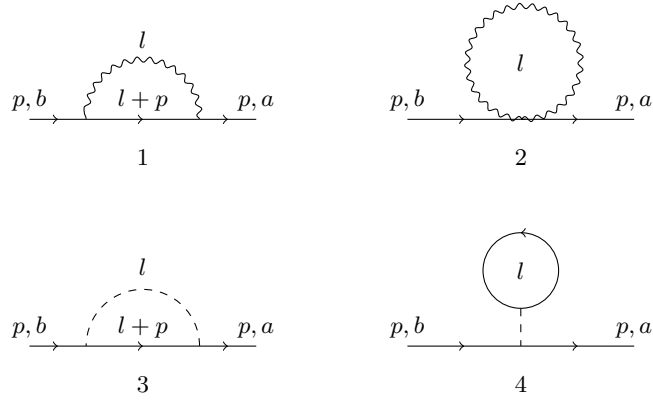

At one-loop level, the spinor self-energy receives contributions from the diagrams shown in Fig.~\ref{fig:self}. Including the corresponding counterterms, the renormalized self-energy is given by
\begin{equation}
-i\Sigma_{ab}(p^2)= -i \Sigma_{ab}^{*}(p^2) + i(p^2-m^2) \delta_{2}\mathds{1}_{ab}- im^2 \delta_m\mathds{1}_{ab},  \label{renoself}
\end{equation}
where $-i\Sigma_{ab}^{*}(p^2)$ denotes the one-loop contribution from
the diagrams shown in Fig.~\ref{fig:self}. Within the on-shell renormalization scheme, the physical self-energy is obtained by requiring the propagator
\begin{equation}
S(p)= \frac{1}{p^2-m^2-\Sigma(p) +i\epsilon}
\end{equation}
to have a simple pole at $p^2=m^2$, leading to the renormalization conditions
\begin{equation}
\Sigma(p^2\rightarrow m^2)=0, \qquad \qquad\left.\frac{\partial \Sigma(p)}{ \partial p^2}\right|_{p^2\rightarrow m^2}=0. \label{con2r}
\end{equation}
These relations fix the counterterms in Eq.~\eqref{renoself} as
\begin{align}
\delta_m =&  -\frac{1}{(4\pi)^2}
\Bigg\{ \left[\frac{1}{\tilde{\epsilon}}
-\ln{\frac{m^2}{\mu^2}}\right]\left[3e^2\left(1+\frac{g^2}{4}\right)+(1-\tau)\lambda_1+\lambda_{2}
+4\vartheta\lambda_{3}  \right]\nonumber\\&\qquad\qquad+e^2\left(7+\frac{g^2}{4}\right)  +(1-\tau)\lambda_1+\lambda_{2}+2\vartheta\lambda_{3} 
\Bigg\},\label{deltam}\\
 \delta_{2}=&
\frac{e^2}{(4\pi)^2}(3-\xi)\left[  \frac{1}{\tilde\epsilon}-\ln{\frac{m^2}{\mu^2}} -  \ln{\frac{m^{2}_{\gamma}}{m^2}}\right],
\label{delta2}
\end{align}
where a small gauge boson mass $m_\gamma$ has been introduced to regulate the infrared divergences. Therefore, the corresponding renormalized spinor self-energy can then be written as
\begin{equation}
\begin{split}
\Sigma(p^2)= \frac{e^2 }{(4\pi)^2} \Bigg\{&
2(p^2+m^{2}) \left[ B_{0}(p^{2},m^{2},m^{2}_{\gamma})-B_0(m^{2},m^{2},m^{2}_{\gamma}) \right] + 4(p^2-m^{2})
 \\
&+2(p^2-m^{2})  \ln{\frac{m^{2}_{\gamma}}{m^2}}+\left(1-\xi\right)\left(p^{2}-m^{2}\right)^2 C_{0}(0,p^{2};m^{2}_{\gamma},m^{2}_{\gamma},m^{2})\Bigg\}.
\end{split}
\end{equation}

\subsection{Spinor-spinor-vector vertex}
\begin{figure}
    \centering
\begin{tikzpicture}
\draw[fermion] (-7,-1.5)  --(-6.5,-1);
\node at (-7,-1.75){$p,b$};
\draw[fermion](-6.5,-1)   --(-5.5,0) ;
\node at (-6.5,-0.25){$l+p$};
\draw[fermion] (-5.5,0)  -- (-4.5,-1) ;
\node at (-4.5,-0.25){$l+p^{\prime}$};
\draw[fermion] (-4.5,-1) --  (-4,-1.5);
\node at (-4,-1.75){$p^{\prime},a$};
\draw [decorate,decoration={snake,segment length=2*3.1416*0.707107,amplitude=1}](-6.5,-1) -- (-4.5,-1);
\node at (-5.5,-1.5) {$l$};
\draw[decorate,decoration={snake,segment length=2*3.1416*0.707107,amplitude=1}] (-5.5,1)--(-5.5,0);
\node at (-5.5 ,1.25) {$q, \mu $};
\node at (-5.5 ,-2.5) {$1$};
\draw[fermion] (-7+5,-1.5)  --(-6.5+5,-1);
\node at (-7+5,-1.75){$p,b$};
\draw[fermion](-6.5+5,-1)   --(-5.5+5,0) ;
\node at (-5.5+5,-0.75){$l+p$};
\draw[fermion] (-5.5+5,0)  -- (-4+5,-1.5);
\draw [decorate,decoration={snake,segment length=2*3.1416*0.707107, amplitude=1}](-5.5+5,0) arc (45:45+180:0.707107);
\node at (-4+5,-1.75){$p^{\prime},a$};
\node at (-6.5+5,.5) {$l$};
\draw[decorate,decoration={snake,segment length=2*3.1416*0.707107,amplitude=1}] (-5.5+5,1)--(-5.5+5,0);
\node at (-5.5+5 ,1.25) {$q, \mu $};
\node at (-5.5+5 ,-2.5) {$2$};
\draw[fermion] (-7+10,-1.5)  --(-5.5+10,0);
\node at (-7+10,-1.75){$p,b$};
\node at (-5.5+10,-0.75){$l+p^{\prime}$};
\draw[fermion] (-5.5+10,0)  -- (-4.5+10,-1);
\draw[fermion] (-4.5+10,-1)  -- (-4+10,-1.5);
\draw [decorate,decoration={snake,segment length=2*3.1416*0.707107, amplitude=1}](-4.5+10,-1) arc (-45:135:0.707107);
\node at (-4+10,-1.75){$p^{\prime},a$};
\node at (-4.5+10,.5) {$l$};
\draw[decorate,decoration={snake,segment length=2*3.1416*0.707107,amplitude=1}] (-5.5+10,1)--(-5.5+10,0);
\node at (-5.5+10 ,1.25) {$q, \mu $};
\node at (-5.5+10 ,-2.5) {$3$};
\draw[fermion] (-7+2.5,-1.5-5)  --(-6.5+2.5,-1-5);
\node at (-7+2.5,-1.75-5){$p,b$};
\draw[fermion](-6.5+2.5,-1-5)   --(-5.5+2.5,0-5) ;
\node at (-6.5+2.5,-0.25-5){$l$};
\draw[fermion] (-5.5+2.5,0-5)  -- (-4.5+2.5,-1-5) ;
\node at (-4.5+2.5,-0.25-5){$l+q$};
\draw[fermion] (-4.5+2.5,-1-5) --  (-4+2.5,-1.5-5);
\node at (-4+2.5,-1.75-5){$p^{\prime},a$};
\draw [scalar](-6.5+2.5,-1-5) -- (-4.5+2.5,-1-5);
\node at (-5.5+2.5,-1.5-5) {};
\draw[decorate,decoration={snake,segment length=2*3.1416*0.707107,amplitude=1}] (-5.5+2.5,1-5)--(-5.5+2.5,0-5);
\node at (-5.5 +2.5,1.25-5) {$q, \mu $};
\node at (-5.5+2.5 ,-2.5-5) {$4$};
\draw[fermion] (-7+7.5,-1.5-5)  --(-5.5+7.5,-1.25-5);
\node at (-7+7.5,-1.75-5){$p,b$};
\node at (-6.5+7.5,-0.25-5){$l+q$};
\node at (-4.5+7.5,-0.25-5){$l$};
\draw[fermion] (-5.5+7.5,-1.25-5) --  (-4+7.5,-1.5-5);
\draw [fermion](-5.5 +7.5,.25-5) arc (90:270:0.5);
\draw [fermion](-5.5 +7.5,-.75-5) arc (-90:90:0.5);
\node at (-4+7.5,-1.75-5){$p^{\prime},a$};
\draw[decorate,decoration={snake,segment length=2*3.1416*0.707107,amplitude=1}] (-5.5+7.5,1-5)--(-5.5+7.5,.25-5);
\draw[scalar] (-5.5+7.5,-1.25-5)--(-5.5+7.5,-.75-5);
\node at (-5.5 +7.5,1.25-5) {$q, \mu $};
\node at (-5.5+7.5 ,-2.5-5) {$5$};
\end{tikzpicture}
    \caption{Feynman diagrams for the spinor-spinor-vector vertex at one-loop level. All external momenta are incoming.}
    \label{fig:ssg}
\end{figure}
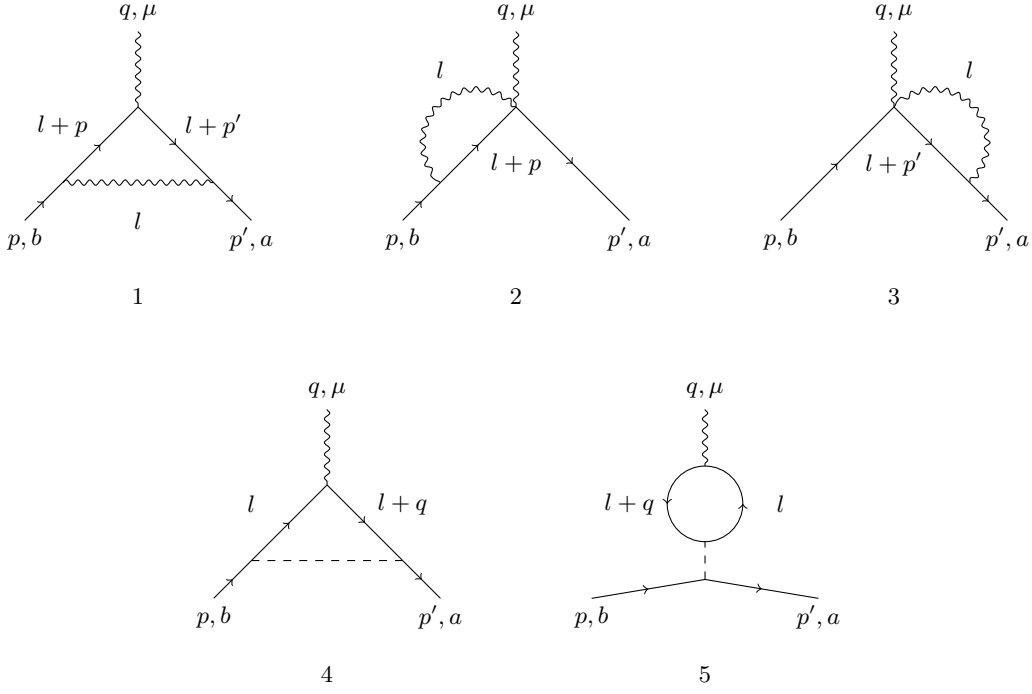
At one-loop order, the spinor-spinor-vector  vertex receives contributions
from the five diagrams shown in Fig.~\ref{fig:ssg}. Including the
corresponding counterterms, the vertex function can be written as
\begin{align}
-ie \Gamma_{ab}^{\mu}(p,p')=-ie V_{ab}^{\mu}(p,p') -ie\Gamma_{ab}^{*\mu}(p,p')
-ie V_{ab}^{\mu}(p,p')\delta_{e}  -e\frac{g}{2}\sigma_{ab}^{\mu\nu}(p'+p)_\nu \delta_g,
\label{gamareno}
\end{align}
 where $-ie\Gamma_{ab}^{*\mu}(p,p')$ denotes the one-loop contribution
from the diagrams shown in Fig.~\ref{fig:ssg}, which satisfies the
Ward--Takahashi identity
\begin{equation}
q^{\mu }\Gamma^{*}_{\mu }(p,p^{\prime})=-\Sigma^{*} (p^{\prime 2})+\Sigma^{*} (p^{2}),
\label{wi32bare}
\end{equation}
or, in its differential form
\begin{equation}
\Gamma^{*}_{\mu }(p,-p)=-\frac{\partial \Sigma^{*} (p^{2})}{\partial p_{\mu}}.
\label{WI32U}
\end{equation}
This constraint implies
\begin{equation}
\delta_e=\delta_{2}= \frac{e^2}{(4\pi)^2}\left(3-\xi\right) \left[ \frac{1}{\tilde{\epsilon}}
-\ln{\frac{m^{2}}{\mu^{2}}}   - \ln{\frac{m_{\gamma}^{2}}{m^{2}}} \right].
\label{deltae}
\end{equation}
Furthermore, using Eqs.~\eqref{ctc} and~\eqref{ctc2}, this relation fixes the counterterm for the spinor-spinor-vector-vector vertex as
\begin{equation}
\delta_3=\delta_{2}.
\label{d3}
\end{equation}
At this point, the on-shell renormalization scheme does not uniquely
fix the counterterm $\delta_g$. We therefore remove the remaining
ultraviolet divergence by choosing
\begin{equation}\label{deltag}
 \delta_{g} =
- \frac{1}{(4\pi)^2} \Bigg\{\left(
\frac{1}{\tilde{\epsilon}}-\ln{\frac{m^{2}}{\mu^{2}}}\right)\left[\left(1-\frac{g^{2}}{4}\right)e^2
 + \lambda_1+\lambda_{2}\right]-2\vartheta\lambda_3\Bigg\},
\end{equation}
such that the on-shell  ($p^2=p'^2=m^2$,  $q^2=(p'-p)^2=m^2_{\gamma}\to 0$) renormalized vertex function becomes
\begin{equation}
-ie \Gamma^{\mu}_{\text{OS}}=-ie\left[(p-p')^\mu \mathds{1}  -i\frac{g}{2}\left(1+\frac{\alpha}{2\pi}\right) \sigma^{\mu\nu} (p'+p)_\nu\right]
 +\frac{ie^2g}{4(4\pi)^2m^{2}} \sigma^{ \alpha\beta} (p-p')_{\alpha} (p+p')_\beta (p-p')^\mu ,
\label{gamarenoos}
\end{equation}
with $\alpha=e^2/(4\pi)$. The tree-level Pauli coupling $g$ therefore receives the same one-loop correction as the one found by Schwinger in the Dirac formalism~\cite{Schwinger:1948iu}. 

\subsection{Spinor-spinor-vector-vector vertex}

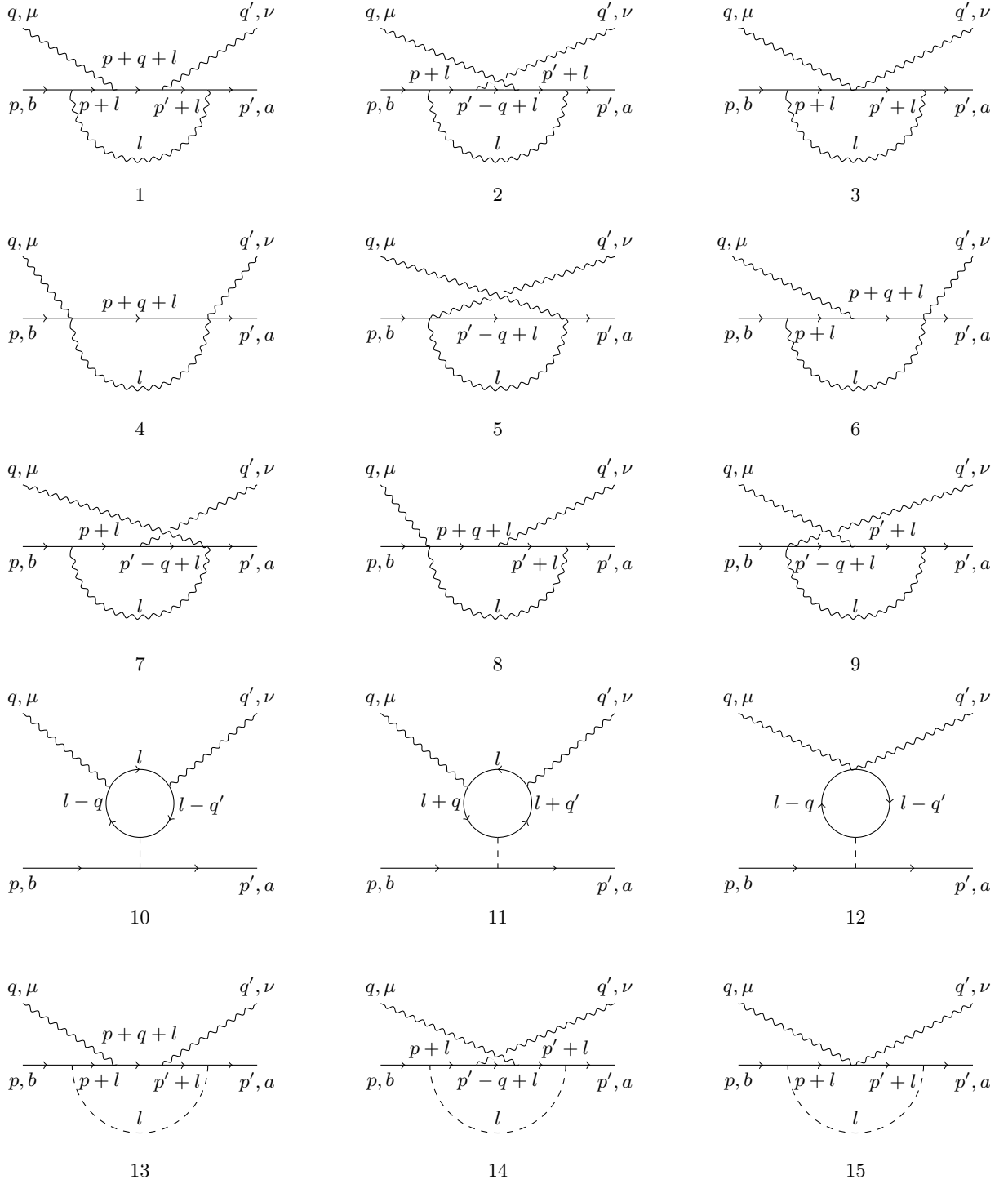
\begin{figure}
    \centering
\begin{tikzpicture}
\draw[fermion] (-6.7-1,-1.5)node[below]{$p,b$}  --(-5.9-1,-1.5) node[below right]{$p+l$};
\draw[fermion] (-3.7-1,-1.5)  --(-2.9-1,-1.5) node[below]{$p',a$};
\draw [decorate,decoration={snake,segment length=5,amplitude=1}](-3.7-1,-1.5) arc (0:-180:1.1);
\node at (-4.8-1,- 3.2) {$1$};
\node at (-4.8-1,- 2.35) {$ l $};
\node at (-4.8-1, -1) {$ p+q+l $};
\draw[fermion] (-5.9-1,-1.5) --(-5.17-1,-1.5);
\draw[fermion] (-5.17-1,-1.5) --(-4.44-1,-1.5) ;
\draw[fermion] (-4.44-1,-1.5) --(-3.7-1,-1.5)node[below left]{$p'+l$} ;
\draw[decorate,decoration={snake,segment length=5,amplitude=1}] (-4.44-1,-1.5) --(-2.9-1,-0.5) node[above]{$q',\nu$};
\draw[decorate,decoration={snake,segment length=5,amplitude=1}] (-5.17-1,-1.5) --(-6.7-1,-0.5) node[above]{$q,\mu$};
\draw[fermion] (-1.9,-1.5) node[below]{$p,b$}  --(-1.1,-1.5)node[above]{$p+l$} ;
\draw[fermion] (1.1,-1.5)  --(1.9,-1.5) node[below]{$p',a$};
\draw [decorate,decoration={snake,segment length=5,amplitude=1}](1.1,-1.5) arc (0:-180:1.1);
\node at (0,- 3.2) {2};
\node at (0,- 2.35) {$ l $};
\node at (0, -1.75) {$ p'-q+l $};
\draw[fermion] (-1.1,-1.5) --(-0.35,-1.5) ;
\draw[fermion] (-0.35,-1.5) --(0.35,-1.5) ;
\draw[fermion] (0.35,-1.5) --(1.1,-1.5) node[above]{$p'+l$};
\draw[decorate,decoration={snake,segment length=5,amplitude=1}] (-0.35,-1.5) --(1.9,-0.5) node[above]{$q' , \nu$};
\draw[draw=white, line width=5.2](0.15 ,-1.4111) --(-1.9,-0.5) ;
\draw[decorate,decoration={snake,segment length=5,amplitude=1}] (0.35 ,-1.5) --(-1.9,-0.5) node[above]{$q,\mu$};
\draw[fermion] (2.9+1,-1.5) node[below]{$p,b$}  --(3.7+1,-1.5) ;
\draw[fermion] (5.9+1,-1.5)  --(6.7+1,-1.5) node[below]{$p',a$};
\draw [decorate,decoration={snake,segment length=5,amplitude=1}](5.9+1,-1.5) arc (0:-180:1.1);
\draw[fermion] (3.7+1,-1.5)node[below right]{$p+l$} --(4.8+1,-1.5) ;
\draw[fermion] (4.8+1,-1.5) --(5.9+1,-1.5)node[below left]{$p'+l$} ;
\draw[decorate,decoration={snake,segment length=5,amplitude=1}] (4.8+1 ,-1.5) --(6.7+1,-0.5) node[above]{$ q',\nu$};
\draw[decorate,decoration={snake,segment length=5,amplitude=1}] (4.8+1 ,-1.5) --(2.9+1,-0.5) node[above]{$q,\mu$};
\node at (4.8+1,- 3.2) {3};
\node at (4.8+1,- 2.35) {$ l $};
\draw[fermion] (-6.7-1,-5.2)node[below]{$p,b$}  --(-5.9-1,-5.2) ;
\draw[fermion] (-3.7-1,-5.2)  --(-2.9-1,-5.2) node[below]{$p',a$};
\draw [decorate,decoration={snake,segment length=5,amplitude=1}](-3.7-1,-5.2) arc (0:-180:1.1);
\node at (-4.8-1,- 7) {4};
\node at (-4.8-1,- 6.15) {$ l $};
\node at (-4.8-1,- 4.95) {$ p+q+l $};
\draw[fermion] (-5.9-1,-5.2) --(-3.7-1,-5.2) ;
\draw[decorate,decoration={snake,segment length=5,amplitude=1}] (-5.9-1,-5.2) --(-6.7-1,-4.2) node[above]{$q,\mu $};
\draw[decorate,decoration={snake,segment length=5,amplitude=1}] (-3.7-1,-5.2) --(-2.9-1,-4.2) node[above]{$q',\nu$};
\draw[fermion] (-1.9,-5.2) node[below]{$p,b$}  --(-1.1,-5.2) ;
\draw[fermion] (1.1,-5.2)  --(1.9,-5.2) node[below]{$p',a$};
\draw [decorate,decoration={snake,segment length=5,amplitude=1}](1.1,-5.2) arc (0:-180:1.1);
\node at (0,- 7) {5};
\node at (0,- 6.15) {$ l $};
\node at (0,-5.45) {$ p'-q+l $};
\draw[fermion] (-1.1,-5.2) --(1.1,-5.2) ;
\draw[decorate,decoration={snake,segment length=5,amplitude=1}] 
(-1.1,-5.2) --(1.9,-4.2) node[above]{$q',\nu$};
\draw[draw=white, line width=5] (.8,-5.1) --(-2.5,-4) ;
\draw[decorate,decoration={snake,segment length=5,amplitude=1}] (1.1,-5.2) --(-1.9,-4.2) node[above]{$q,\mu $};
\draw[fermion] (2.9+1,-5.2) node[below]{$p,b$}  --(3.7+1,-5.2) ;
\draw[fermion] (5.9+1,-5.2)  --(6.7+1,-5.2) node[below]{$p',a$};
\draw [decorate,decoration={snake,segment length=5,amplitude=1}](5.9+1,-5.2) arc (0:-180:1.1);
\node at (4.8+1,- 7) {$ 6 $};
\node at (4.8+1,- 6.15) {$ l $};
\draw[fermion] (3.7+1,-5.2)node[below right]{$p+l$} --(4.8+1,-5.2) ;
\draw[fermion] (4.8+1,-5.2) --(5.9+1,-5.2);
\node at (5.3+1,-4.8){$p+q+l$};
\draw[decorate,decoration={snake,segment length=5,amplitude=1}] (4.8+1,-5.2) --(2.8+1,-4.2) node[above]{$q,\mu $};
\draw[decorate,decoration={snake,segment length=5,amplitude=1}] (5.9+1,-5.2) --(6.7+1,-4.2)node[above]{$q',\nu$};
\draw[fermion] (-6.7-1,-8.9)node[below]{$p,b$}  --(-5.9-1,-8.9) ;
\draw[fermion] (-3.7-1,-8.9)  --(-2.9-1,-8.9) node[below]{$p',a$};
\draw [decorate,decoration={snake,segment length=5,amplitude=1}](-3.7-1,-8.9) arc (0:-180:1.1);
\node at (-4.8-1,- 10.8) {7};
\node at (-4.8-1,- 9.85) {$ l $};
\draw[fermion] (-5.9-1,-8.9)node[above right]{$p+l$} --(-4.8-1,-8.9) ;
\draw[fermion] (-4.8-1,-8.9) -- (-3.7-1,-8.9) node[below left]{$p'-q+l$};
\draw[decorate,decoration={snake,segment length=5,amplitude=1}] (-4.8-1,-8.9) --(-2.9-1,-7.9) node[above]{$q',\nu $};
\draw[draw=white, line width=5] (-4.2-1,-8.73333) --(-5-1,-8.46667);
\draw[decorate,decoration={snake,segment length=5,amplitude=1}] (-3.7-1,-8.9) --(-6.7-1,-7.9) node[above]{$q,\mu$};
\draw[fermion] (-1.9,-8.9) node[below]{$p,b$}  --(-1.1,-8.9)node[above right]{$p+q+l$} ;
\draw[fermion]  (1.1,-8.9)  --(1.9,-8.9) node[below]{$p',a$};
\draw [decorate,decoration={snake,segment length=5,amplitude=1}](1.1,-8.9) arc (0:-180:1.1);
\node at (0,- 10.8) {8};
\node at (0,- 9.85) {$ l $};
\draw[fermion] (-1.1,-8.9) --(0,-8.9) ;
\draw[fermion] (0,-8.9)--(1.1,-8.9)node[below left]{$p'+l$}  ;
\draw[decorate,decoration={snake,segment length=5,amplitude=1}] (0,-8.9) --(1.9,-7.9) node[above]{$q',\nu $};
\draw[decorate,decoration={snake,segment length=5,amplitude=1}] (-1.1,-8.9) --(-1.9,-7.9) node[above]{$q,\mu$};
\draw[fermion] (2.9+1,-8.9) node[below]{$p,b$}  --(3.7+1,-8.9) node[below right]{$p'-q+l$};
\draw[fermion] (5.9+1,-8.9)  --(6.7+1,-8.9) node[below]{$p',a$};
\draw [decorate,decoration={snake,segment length=5,amplitude=1}](5.9+1,-8.9) arc (0:-180:1.1);
\node at (4.8+1,- 10.8) {9};
\node at (4.8+1,- 9.85) {$ l $};
\draw[fermion] (3.7+1,-8.9)--(4.8+1,-8.9) ;
\draw[fermion] (4.8+1,-8.9)--(5.9+1,-8.9) node[above left]{$p'+l$};
\draw[decorate,decoration={snake,segment length=5,amplitude=1}] (3.7+1,-8.9) --(6.7+1,-7.9) node[above]{$q',\nu $};
\draw[draw=white, line width=5] (4.5+1,-8.74211) --(2.9+1,-7.9);
\draw[decorate,decoration={snake,segment length=5,amplitude=1}] (4.8+1,-8.9) --(2.9+1,-7.9) node[above]{$q, \mu$};
\draw[fermion] (-6.7-1,-13.61-.5) node[below]{$p,b$} --(-4.8-1,-13.61-.5) ;
\draw[fermion] (-4.8-1,-13.61-.5)  --(-2.9-1,-13.61-.5)node[below]{$p',a$} ;
\draw[scalar] (-4.8-1,-13.61)--(-4.8-1,-13.61-.5);
\draw[decorate,decoration={snake,segment length=5,amplitude=1}] (-5.28-1 ,-12.785) --(-6.7-1,-11.6) node[above]{$q, \mu$};
\draw[decorate,decoration={snake,segment length=5,amplitude=1}] (-4.324-1 ,-12.785) --(-2.9-1,-11.6) node[above]{$q',\nu $};
\draw [fermion](-4.8-1,-13.61) arc (270: 150:0.55) node[below left]{$l-q$};
\draw [fermion](-5.28-1,-12.785) arc (150: 30:0.55)  ;
\draw [fermion](-4.324-1,-12.785)  arc (30: -90:0.55) ;
\node at (-3.8-1,-13.085) {$l-q'$};
\node at (-4.8-1,-12.3) {$l$};
\node at (-4.8-1,-13.61-1.3) {$10$};
\draw[fermion] (-1.9,-13.61-.5) node[below]{$p,b$} --(0,-13.61-.5) ;
\draw[fermion] (0,-13.61-.5)  --(1.9,-13.61-.5) node[below]{$p',a$} ;
\draw[scalar] (0,-13.61)--(0,-13.61-.5);
\draw[decorate,decoration={snake,segment length=5,amplitude=1}] (-0.48 ,-12.785) --(-1.9,-11.6) node[above]{$q, \mu$};
\draw[decorate,decoration={snake,segment length=5,amplitude=1}] (0.476 ,-12.785) --(1.9,-11.6)node[above]{$q',\nu $} ;
\draw [fermion](0,-13.61) arc (-90: 30:0.55)node[below right]{$l+q'$};
\draw [fermion](-0.48,-12.785) node[below left]{$l+q$} arc (150: 270:0.55);
\draw [fermion](0.476,-12.785)  arc (30: 150:0.55);
\node at  (0, -12.33) {$l$} ;
\node at (0,-13.61-1.3) {$11$};
\draw[fermion] (2.9+1,-13.61-.5)node[below]{$p,b$}  --(4.8+1,-13.61-.5) ;
\draw[fermion] (4.8+1 ,-13.61-.5)  --(6.7+1,-13.61-.5) node[below]{$p',a$} ;
\draw[scalar] (4.8+1,-13.61)--(4.8+1,-13.61-.5);
\draw[decorate,decoration={snake,segment length=5,amplitude=1}] (4.8+1,-12.51) --(6.7+1, -11.6) node[above]{$q',\nu $} ;
\draw[decorate,decoration={snake,segment length=5,amplitude=1}] (4.8+1 ,-12.51) --(2.9+1, -11.6) node[above]{$q, \mu$};
\draw [fermion](4.8+1,-13.61)  arc (270: 90:0.55);
\draw [fermion](4.8+1,-12.51) arc (90: -90:0.55);
\node at (3.8+1, -13.06){$l-q$};
\node at (5.9+1,-13.06){$l-q'$};
\node at (4.8+1,-13.61-1.3) {$12$};
\draw[fermion] (-6.7-1,-1.5-15.8)node[below]{$p,b$}  --(-5.9-1,-1.5-15.8) node[below right]{$p+l$};
\draw[fermion] (-3.7-1,-1.5-15.8)  --(-2.9-1,-1.5-15.8) node[below]{$p',a$};
\draw [scalar](-3.7-1,-1.5-15.8) arc (0:-180:1.1);
\node at (-4.8-1,- 3.2-15.8) {13};
\node at (-4.8-1,- 2.35-15.8) {$ l $};
\node at (-4.8-1, -1-15.8) {$ p+q+l $};
\draw[fermion] (-5.9-1,-1.5-15.8) --(-5.17-1,-1.5-15.8);
\draw[fermion] (-5.17-1,-1.5-15.8) --(-4.44-1,-1.5-15.8) ;
\draw[fermion] (-4.44-1,-1.5-15.8) --(-3.7-1,-1.5-15.8)node[below left]{$p'+l$} ;
\draw[decorate,decoration={snake,segment length=5,amplitude=1}] (-4.44-1,-1.5-15.8) --(-2.9-1,-0.5-15.8) node[above]{$q',\nu$};
\draw[decorate,decoration={snake,segment length=5,amplitude=1}] (-5.17-1,-1.5-15.8) --(-6.7-1,-0.5-15.8) node[above]{$q,\mu$};
\draw[fermion] (-1.9,-1.5-15.8) node[below]{$p,b$}  --(-1.1,-1.5-15.8)node[above]{$p+l$} ;
\draw[fermion] (1.1,-1.5-15.8)  --(1.9,-1.5-15.8) node[below]{$p',a$};
\draw [scalar](1.1,-1.5-15.8) arc (0:-180:1.1);
\node at (0,- 3.2-15.8) {14};
\node at (0,- 2.35-15.8) {$ l $};
\node at (0, -1.75-15.8) {$ p'-q+l $};
\draw[fermion] (-1.1,-1.5-15.8) --(-0.35,-1.5-15.8) ;
\draw[fermion] (-0.35,-1.5-15.8) --(0.35,-1.5-15.8) ;
\draw[fermion] (0.35,-1.5-15.8) --(1.1,-1.5-15.8) node[above]{$p'+l$};
\draw[decorate,decoration={snake,segment length=5,amplitude=1}] (-0.35,-1.5-15.8) --(1.9,-0.5-15.8) node[above]{$q' , \nu$};
\draw[draw=white, line width=5.2](0.15 ,-1.4111-15.8) --(-1.9,-0.5-15.8) ;
\draw[decorate,decoration={snake,segment length=5,amplitude=1}] (0.35 ,-1.5-15.8) --(-1.9,-0.5-15.8) node[above]{$q,\mu$};
\draw[fermion] (2.9+1,-1.5-15.8) node[below]{$p,b$}  --(3.7+1,-1.5-15.8) ;
\draw[fermion] (5.9+1,-1.5-15.8)  --(6.7+1,-1.5-15.8) node[below]{$p',a$};
\draw [scalar](5.9+1,-1.5-15.8) arc (0:-180:1.1);
\draw[fermion] (3.7+1,-1.5-15.8)node[below right]{$p+l$} --(4.8+1,-1.5-15.8) ;
\draw[fermion] (4.8+1,-1.5-15.8) --(5.9+1,-1.5-15.8)node[below left]{$p'+l$} ;
\draw[decorate,decoration={snake,segment length=5,amplitude=1}] (4.8+1 ,-1.5-15.8) --(2.9+1,-0.5-15.8) node[above]{$q,\mu$};
\draw[decorate,decoration={snake,segment length=5,amplitude=1}] (4.8+1 ,-1.5-15.8) --(6.7+1,-0.5-15.8) node[above]{$ q',\nu$};
\node at (4.8+1,- 2.35-15.8) {$ l $};
\node at (4.8+1,- 3.2-15.8) {15};
\end{tikzpicture}
    \caption{Feynman diagrams for the spinor-spinor-vector-vector vertex at one-loop level. All external momenta are incoming.}
    \label{fig:ssgg}
\end{figure}

The spinor-spinor-vector-vector vertex function at one-loop level is
\begin{equation}
ie^2 \Gamma_{ab}^{\mu\nu}(p,q,p',q') = ie^2 V_{ab}^{\mu\nu}+ie^2 \Gamma_{ab}^{*\mu\nu}(p,q,p',q') + ie^2 V_{ab}^{\mu\nu}\delta_{3},
\label{segwar1}
\end{equation}
where the one-loop corrections $ie^2 \Gamma_{ab}^{*\mu\nu}(p,q,p',q')$ are given by the diagrams of Fig.~\ref{fig:ssgg}.
The counterterm $\delta_{3}$ has already been fixed in Eqs.~\eqref{delta2} and~\eqref{d3}. An explicit calculation shows that this counterterm renders the one-loop vertex finite.

\subsection{Four-vector vertex}

\begin{figure}
    \centering
\begin{tikzpicture}
\draw[decorate,decoration={snake,segment length=5,amplitude=1}](-6.7,-2.6)  --(-5.117,-2.19952);
\node at  (-6.7, -2.8){$ q_{2}, \nu $} ;
\draw[decorate,decoration={snake,segment length=5,amplitude=1}](-6.7,-0.5) -- (-5.867,-0.900481);
\node at (-6.7,-0.3) {$ q_{1}, \mu  $};
\draw[fermion] (-5.867,-0.900481)--(-4.367,-0.900481);
\draw[fermion] (-4.367,-0.900481) --(-5.117,-2.19952);
\draw[fermion] (-5.117,-2.19952)-- (-5.867,-0.900481);
\draw[decorate,decoration={snake,segment length=5,amplitude=1}] (-4.367,-0.900481) --(-3.54,-0.5) ;
\node at (-3.54,-0.3) {$ q'_{1}, \alpha $};
\draw[decorate,decoration={snake,segment length=5,amplitude=1}] (-5.117,-2.19952) --(-3.54,-2.6) ;
\node at (-3.54,-2.8){$ q'_{2}, \beta $};
\node at (-5.117,-3){$1$};
\draw[decorate,decoration={snake,segment length=5,amplitude=1}](-6.7+5,-2.6)  --(-5.117+5,-2.19952);
\node at  (-6.7+5, -2.8){$ q_{2}, \nu $} ;
\draw[decorate,decoration={snake,segment length=5,amplitude=1}](-6.7+5,-0.5) -- (-5.867+5,-0.900481);
\node at (-6.7+5,-0.3) {$ q_{1}, \mu  $};
\draw[fermion] (-5.867+5,-0.900481)--(-4.367+5,-0.900481);
\draw[fermion] (-4.367+5,-0.900481) --(-5.117+5,-2.19952);
\draw[fermion] (-5.117+5,-2.19952)-- (-5.867+5,-0.900481);
\draw[decorate,decoration={snake,segment length=5,amplitude=1}](-4.367+5,-0.900481)  --(-3.54+5,-0.5) ;
\node at (-3.54+5,-0.3) {$ q'_{2}, \beta $};
\draw[decorate,decoration={snake,segment length=5,amplitude=1}]  (-5.117+5,-2.19952) --(-3.54+5,-2.6) ;
\node at (-3.54+5,-2.8){$ q'_{1}, \alpha $};
\node at (-5.117+5,-3){$2$};
\draw[decorate,decoration={snake,segment length=5,amplitude=1}](-6.7+10,-2.6)  --(-5.55+10,-2.3);
\node at  (-6.7+10, -2.8){$ q_{2}, \nu $} ;
\draw[decorate,decoration={snake,segment length=5,amplitude=1}](-6.7+10,-0.5) -- (-5.55+10,-0.8);
\node at (-6.7+10,-0.3) {$ q_{1}, \mu  $};
\draw[fermion] (-5.55+10,-2.3)--(-5.55+10,-0.8);
\draw[fermion] (-5.55+10,-0.8) --(-4.25+10, -1.55 );
\draw[fermion] (-4.25+10, -1.55 )-- (-5.55+10,-2.3);
\draw[decorate,decoration={snake,segment length=5,amplitude=1}] (-4.25+10, -1.55 ) --(-3.54+10,-0.5) ;
\node at (-3.54+10,-0.3) {$ q'_{1}, \alpha $};
\draw[decorate,decoration={snake,segment length=5,amplitude=1}] (-4.25+10, -1.55 )  --(-3.54+10,-2.6) ;
\node at (-3.54+10,-2.8){$ q'_{2}, \beta $};
\node at (-5.117+10,-3){$3$};
\draw[decorate,decoration={snake,segment length=5,amplitude=1}](-6.7,-2.6-4)  --(-5.867,-2.19952-4);
\node at  (-6.7, -2.8-4){$ q_{2}, \nu $} ;
\draw[decorate,decoration={snake,segment length=5,amplitude=1}](-6.7,-0.5-4) -- (-5.117,-0.900481-4);
\node at (-6.7,-0.3-4) {$ q_{1}, \mu  $};
\draw[fermion] (-4.367,-2.19952-4)--(-5.867,-2.19952-4);
\draw[fermion] (-5.867,-2.19952-4) --(-5.117,-0.900481-4);
\draw[fermion] (-5.117,-0.900481-4)-- (-4.367,-2.19952-4);
\draw[decorate,decoration={snake,segment length=5,amplitude=1}] (-5.117,-0.900481-4) --(-3.54,-0.5-4) ;
\node at (-3.54,-0.3-4) {$ q'_{1}, \alpha $};
\draw[decorate,decoration={snake,segment length=5,amplitude=1}] (-4.367,-2.19952-4)  --(-3.54,-2.6-4) ;
\node at (-3.54,-2.8-4){$ q'_{2}, \beta $};
\node at (-5.117,-3-4){$4$};
\draw[decorate,decoration={snake,segment length=5,amplitude=1}](-6.7+5,-2.6-4)  --(-5.867+5,-2.19952-4);
\node at  (-6.7+5, -2.8-4){$ q_{2}, \nu $} ;
\draw[decorate,decoration={snake,segment length=5,amplitude=1}](-6.7+5,-0.5-4) -- (-5.117+5,-0.900481-4);
\node at (-6.7+5,-0.3-4) {$ q_{1}, \mu  $};
\draw[fermion] (-4.367+5,-2.19952-4)--(-5.867+5,-2.19952-4);
\draw[fermion] (-5.867+5,-2.19952-4) --(-5.117+5,-0.900481-4);
\draw[fermion] (-5.117+5,-0.900481-4)-- (-4.367+5,-2.19952-4);
\draw[decorate,decoration={snake,segment length=5,amplitude=1}] (-5.117+5,-0.900481-4) --(-3.54+5,-0.5-4) ;
\node at (-3.54+5,-0.3-4){$ q'_{2}, \beta $};
\draw[decorate,decoration={snake,segment length=5,amplitude=1}] (-4.367+5,-2.19952-4)  --(-3.54+5,-2.6-4) ;
\node at (-3.54+5,-2.8-4){$ q'_{1}, \alpha $};
\node at (-5.117+5,-3-4){$5$};
\draw[decorate,decoration={snake,segment length=5,amplitude=1}](-6.7+10,-2.6-4)  --(-5.98+10,-1.55-4);
\node at  (-6.7+10, -2.8-4){$ q_{2}, \nu $} ;
\draw[decorate,decoration={snake,segment length=5,amplitude=1}](-6.7+10,-0.5-4) -- (-5.98+10,-1.55 -4);
\node at (-6.7+10,-0.3-4) {$ q_{1}, \mu  $};
\draw[fermion] (-5.98+10, -1.55-4)--(-4.684+10, -0.8-4);
\draw[fermion] (-4.684+10, -0.8-4) --(-4.684+10,-2.3-4);
\draw[fermion] (-4.684+10, -2.3 -4)-- (-5.98+10,-1.55 -4);
\draw[decorate,decoration={snake,segment length=5,amplitude=1}] (-4.684+10, -0.8-4) --(-3.54+10,-0.5-4) ;
\node at (-3.54+10,-0.3-4) {$ q'_{1}, \alpha $};
\draw[decorate,decoration={snake,segment length=5,amplitude=1}] (-4.684+10,-2.3-4)  --(-3.54+10,-2.6-4) ;
\node at (-3.54+10,-2.8-4){$ q'_{2}, \beta $};
\node at (-5.117+10,-3-4){$6$};
\draw[decorate,decoration={snake,segment length=5,amplitude=1}](-6.7,-2.6-8)  --(-5.717,-2.15-8);
\node at  (-6.7, -2.8-8){$ q_{2}, \nu $} ;
\draw[decorate,decoration={snake,segment length=5,amplitude=1}](-6.7,-0.5-8) -- (-5.717,-0.95-8);
\node at (-6.7,-0.3-8) {$ q_{1}, \mu  $};
\draw[fermion] (-5.717,-2.15-8) --(-5.717,-0.95-8);
\draw[fermion] (-5.717,-0.95-8) --(-4.517,-0.95-8);
\draw[fermion] (-4.517,-0.95-8)-- (-4.517,-2.15-8);
\draw[fermion] (-4.517,-2.15-8)-- (-5.717,-2.15-8);
\draw[decorate,decoration={snake,segment length=5,amplitude=1}] (-4.517,-0.95-8) --(-3.54,-0.5-8) ;
\node at (-3.54,-0.3-8) {$ q'_{1}, \alpha $};
\draw[decorate,decoration={snake,segment length=5,amplitude=1}] (-4.517,-2.15-8) --(-3.54,-2.6-8) ;
\node at (-3.54,-2.8-8){$ q'_{2}, \beta $};
\node at (-5.117,-3-8){$7$};
\draw[decorate,decoration={snake,segment length=5,amplitude=1}](-6.7+5,-2.6-8)  --(-5.717+5,-2.15-8);
\node at  (-6.7+5, -2.8-8){$ q_{2}, \nu $} ;
\draw[decorate,decoration={snake,segment length=5,amplitude=1}](-6.7+5,-0.5-8) -- (-5.717+5,-0.95-8);
\node at (-6.7+5,-0.3-8) {$ q_{1}, \mu  $};
\draw[fermion] (-5.717+5,-2.15-8) --(-5.717+5,-0.95-8);
\draw[fermion] (-5.717+5,-0.95-8) --(-4.517+5,-0.95-8);
\draw[fermion] (-4.517+5,-0.95-8)-- (-4.517+5,-2.15-8);
\draw[fermion] (-4.517+5,-2.15-8)-- (-5.717+5,-2.15-8);
\draw[decorate,decoration={snake,segment length=5,amplitude=1}] (-4.517+5,-0.95-8) --(-3.54+5,-0.5-8) ;
\node at (-3.54+5,-0.3-8) {$ q'_{2}, \beta $};
\draw[decorate,decoration={snake,segment length=5,amplitude=1}] (-4.517+5,-2.15-8) --(-3.54+5,-2.6-8) ;
\node at (-3.54+5,-2.8-8){$ q'_{1}, \alpha $};
\node at (-5.117+5,-3-8){$8$};
\draw[decorate,decoration={snake,segment length=5,amplitude=1}](-6.7+10,-2.6-8)  --(-5.717+10,-2.15-8);
\node at  (-6.7+10, -2.8-8){$ q_{2}, \nu $} ;
\draw[decorate,decoration={snake,segment length=5,amplitude=1}](-6.7+10,-0.5-8) -- (-5.717+10,-0.95-8);
\node at (-6.7+10,-0.3-8) {$ q_{1}, \mu  $};
\draw[fermion] (-4.517+10,-0.95-8)--(-5.717+10,-0.95-8) ;
\draw[fermion] (-5.717+10,-2.15-8)--(-5.117+10,-1.55-8);
\draw[fermion] (-5.117+10,-1.55-8)--(-4.517+10,-0.95-8);
\draw[fermion] (-4.517+10,-2.15-8)-- (-5.717+10,-2.15-8);
\draw[draw=white,line width=5] (-5.5+10,-1.167-8)--(-5+10,-1.667-8);
\draw[fermion] (-5.717+10,-0.95-8)--(-4.517+10,-2.15-8);
\draw[decorate,decoration={snake,segment length=5,amplitude=1}] (-4.517+10,-0.95-8) --(-3.54+10,-0.5-8) ;
\node at (-3.54+10,-0.3-8) {$ q'_{1}, \alpha $};
\draw[decorate,decoration={snake,segment length=5,amplitude=1}] (-4.517+10,-2.15-8) --(-3.54+10,-2.6-8) ;
\node at (-3.54+10,-2.8-8){$ q'_{2}, \beta $};
\node at (-5.117+10,-3-8){$9$};
\draw[decorate,decoration={snake,segment length=5,amplitude=1}](-6.7,-2.6-12)  --(-5.117-0.7, -1.55-12);
\node at  (-6.7, -2.8-12){$ q_{2}, \nu $} ;
\draw[decorate,decoration={snake,segment length=5,amplitude=1}](-6.7,-0.5-12) -- (-5.117-0.7, -1.55-12);
\node at (-6.7,-0.3-12) {$ q_{1}, \mu  $};
\draw [fermion](-5.117-0.7, -1.55-12)  arc (180: 0:0.7);
\draw[fermion] (-5.117+0.7, -1.55-12) arc (0:-180:0.7);
\draw[decorate,decoration={snake,segment length=5,amplitude=1}](-5.117+0.7, -1.55-12) --(-3.54,-0.5-12) ;
\node at (-3.54,-0.3-12) {$ q'_{1}, \alpha $};
\draw[decorate,decoration={snake,segment length=5,amplitude=1}] (-5.117+0.7, -1.55-12) --(-3.54,-2.6-12) ;
\node at (-3.54,-2.8-12){$ q'_{2}, \beta $};
\node at (-5.117,-3-12){$10$};
\draw[decorate,decoration={snake,segment length=5,amplitude=1}](-6.7+5,-2.6-12)  --(-5.117+5, -1.55-0.7-12);
\node at  (-6.7+5, -2.8-12){$ q_{2}, \nu $} ;
\draw[decorate,decoration={snake,segment length=5,amplitude=1}](-6.7+5,-0.5-12) -- (-5.117+5, -1.55+0.7-12);
\node at (-6.7+5,-0.3-12) {$ q_{1}, \mu  $};
\draw [fermion](-5.117+5, -1.55+0.7-12)  arc (90: -90:0.7);
\draw[fermion] (-5.117+5, -1.55-0.7-12) arc (270:90:0.7);
\draw[decorate,decoration={snake,segment length=5,amplitude=1}](-5.117+5, -1.55+0.7-12)--(-3.54+5,-0.5-12) ;
\node at (-3.54+5,-0.3-12) {$ q'_{1}, \alpha $};
\draw[decorate,decoration={snake,segment length=5,amplitude=1}] (-5.117+5, -1.55-0.7-12) --(-3.54+5,-2.6-12) ;
\node at (-3.54+5,-2.8-12){$ q'_{2}, \beta $};
\node at (-5.117+5,-3-12){$11$};
\draw[decorate,decoration={snake,segment length=5,amplitude=1}](-6.7+10,-2.6-12)  --(-5.117+10, -1.55-0.7-12);
\node at  (-6.7+10, -2.8-12){$ q_{2}, \nu $} ;
\draw[decorate,decoration={snake,segment length=5,amplitude=1}](-6.7+10,-0.5-12) -- (-5.117+10, -1.55+0.7-12);
\node at (-6.7+10,-0.3-12) {$ q_{1}, \mu  $};
\draw [fermion](-5.117+10, -1.55+0.7-12)  arc (90: -90:0.7);
\draw[fermion] (-5.117+10, -1.55-0.7-12) arc (270:90:0.7);
\draw[decorate,decoration={snake,segment length=5,amplitude=1}](-5.117+10, -1.55+0.7-12)--(-3.54+10,-0.5-12) ;
\node at (-3.54+10,-0.3-12){$ q'_{2}, \beta $};
\draw[decorate,decoration={snake,segment length=5,amplitude=1}] (-5.117+10, -1.55-0.7-12) --(-3.54+10,-2.6-12) ;
\node at (-3.54+10,-2.8-12){$ q'_{1}, \alpha $};
\node at (-5.117+10,-3-12){$12$};
\end{tikzpicture}
    \caption{Feynman diagrams for the four-vector vertex at one-loop level. All external momenta are incoming.}
    \label{fig:gggg}
\end{figure}
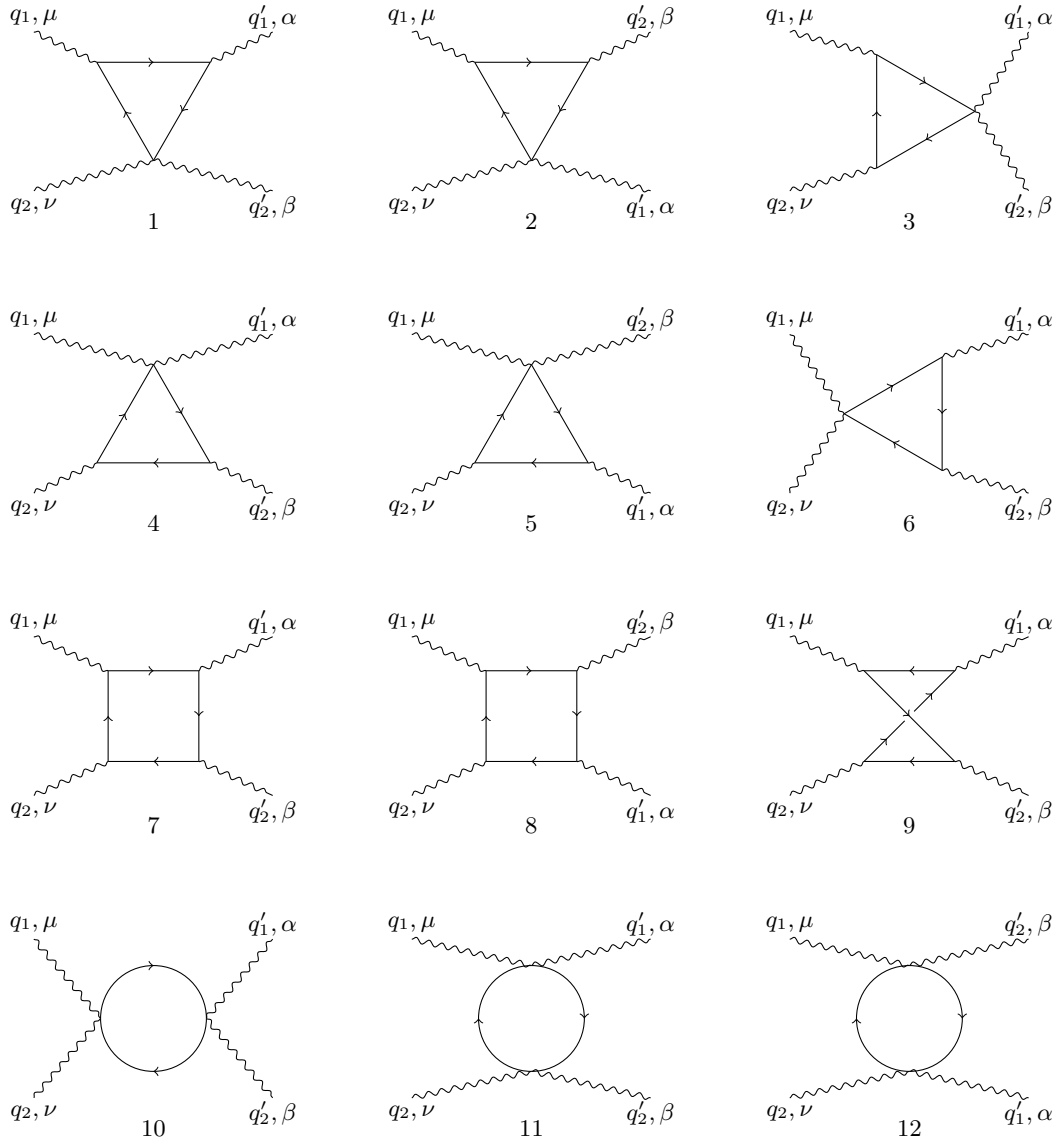
The four-vector vertex is absent at tree-level and is generated at one-loop level by the diagrams shown in Fig.~\ref{fig:gggg}. Direct calculation shows that the resulting amplitude is finite, as expected. Similarly, the one-loop three-vector vertex function vanishes identically by charge-conjugation symmetry.

\subsection{Four-spinor vertex}

\begin{figure}
    \centering
\begin{tikzpicture}
\draw[fermion](-6.7,-2.6)  --(-5.717,-2.15);
\node at  (-6.7, -2.9){$ p_{2}, d $} ;
\draw[fermion](-6.7,-0.5) -- (-5.717,-0.95);
\node at (-6.7,-0.2) {$ p_{1}, b $};
\draw[decorate,decoration={snake,segment length=5,amplitude=1}] (-5.717,-2.15) --(-5.717,-0.95);
\draw[fermion] (-5.717,-0.95) --(-4.517,-0.95);
\draw[decorate,decoration={snake,segment length=5,amplitude=1}] (-4.517,-0.95)-- (-4.517,-2.15);
\draw[fermion] (-5.717,-2.15)-- (-4.517,-2.15);
\draw[fermion] (-4.517,-0.95) --(-3.54,-0.5) ;
\node at (-3.54,-0.2) {$ p'_{1}, a $};
\draw[fermion] (-4.517,-2.15) --(-3.54,-2.6) ;
\node at (-3.54,-2.9){$ p'_{2}, c $};
\node at (-5.117,-3.2){$1$};
\draw[fermion](-6.7+5,-2.6)  --(-5.717+5,-2.15);
\node at  (-6.7+5, -2.9){$ p_{2}, d $} ;
\draw[fermion](-6.7+5,-0.5) --(-5.717+5,-0.95);
\node at (-6.7+5,-0.2) {$ p_{1}, b $};
\draw[decorate,decoration={snake,segment length=5,amplitude=1}] (-5.717+5,-2.15)-- (-4.517+5,-0.95);
\draw[fermion] (-5.717+5,-0.95)--(-4.517+5,-0.95)  ;
\draw[draw=white,line width=4.5] (-5.4+5,-1.267) -- (-4.7+5,-1.967);
\draw[decorate,decoration={snake,segment length=5,amplitude=1}] (-5.717+5,-0.95) -- (-4.517+5,-2.15);
\draw[fermion] (-5.717+5,-2.15)-- (-4.517+5,-2.15);
\draw[fermion] (-4.517+5,-0.95) --(-3.54+5,-0.5) ;
\node at (-3.54+5,-0.2) {$ p'_{1}, a $};
\draw[fermion] (-4.517+5,-2.15) --(-3.54+5,-2.6) ;
\node at (-3.54+5,-2.9){$ p'_{2}, c $};
\node at (-5.117+5,-3.2){$2$};
\draw[fermion](-6.7+10,-2.6)  --(-5.117+10,-2.19952);
\node at  (-6.7+10, -2.9){$ p_{2}, d $} ;
\draw[fermion](-6.7+10,-0.5) -- (-5.867+10,-0.900481);
\node at (-6.7+10,-0.2) {$ p_{1}, b $};
\draw[fermion] (-5.867+10,-0.900481)--(-4.367+10,-0.900481);
\draw[decorate,decoration={snake,segment length=5,amplitude=1}] (-4.367+10,-0.900481) --(-5.117+10,-2.19952);
\draw[decorate,decoration={snake,segment length=5,amplitude=1}] (-5.117+10,-2.19952)-- (-5.867+10,-0.900481);
\draw[fermion] (-4.367+10,-0.900481) --(-3.54+10,-0.5) ;
\node at (-3.54+10,-0.2) {$  p'_{1},a $};
\draw[fermion] (-5.117+10,-2.19952) --(-3.54+10,-2.6) ;
\node at (-3.54+10,-2.9){$  p'_{2}, c $};
\node at (-5.117+10,-3.2){$3$};
\draw[fermion](-6.7+15,-2.6)  --(-5.867+15,-2.19952);
\node at  (-6.7+15, -2.9){$ p_{2},d  $} ;
\draw[fermion](-6.7+15,-0.5) -- (-5.117+15,-0.900481);
\node at (-6.7+15,-0.2) {$  p_{1}, b  $};
\draw[fermion] (-5.867+15,-2.19952)--(-4.367+15,-2.19952);
\draw[decorate,decoration={snake,segment length=5,amplitude=1}] (-5.867+15,-2.19952) --(-5.117+15,-0.900481);
\draw[decorate,decoration={snake,segment length=5,amplitude=1}] (-5.117+15,-0.900481)-- (-4.367+15,-2.19952);
\draw[fermion] (-5.117+15,-0.900481) --(-3.54+15,-0.5) ;
\node at (-3.54+15,-0.2) {$  p'_{1},a $};
\draw[fermion] (-4.367+15,-2.19952)  --(-3.54+15,-2.6) ;
\node at (-3.54+15,-2.9){$ p'_{2}, c  $};
\node at (-5.117+15,-3.2){$4$};
\draw[fermion](-6.7,-2.6-4)  --(-5.117, -1.55-0.7-4);
\node at  (-6.7, -2.9-4){$ p_{2}, d $} ;
\draw[fermion](-6.7,-0.5-4) -- (-5.117, -1.55+0.7-4);
\node at (-6.7,-0.2-4) {$ p_{1}, b $};
\draw[decorate,decoration={snake,segment length=2*3.1416*.7,amplitude=1}](-5.117, -1.55+0.7-4)  arc (90: 90+360:0.7);
\draw[fermion](-5.117, -1.55+0.7-4) --(-3.54,-0.5-4) ;
\node at (-3.54,-0.2-4) {$  p'_{1},a$};
\draw[fermion] (-5.117, -1.55-0.7-4) --(-3.54,-2.6-4) ;
\node at (-3.54,-2.9-4){$  p'_{2}, c  $};
\node at (-5.117,-3.2-4){$5$};
\draw[fermion](-6.7+5,-2.6-4)  --(-5.717+5,-2.15-4);
\node at  (-6.7+5, -2.9-4){$ p_{2}, d $} ;
\draw[fermion](-6.7+5,-0.5-4) -- (-5.717+5,-0.95-4);
\node at (-6.7+5,-0.2-4) {$ p_{1}, b $};
\draw[decorate,decoration={snake,segment length=5,amplitude=1}] (-5.717+5,-2.15-4) --(-5.717+5,-0.95-4);
\draw[fermion] (-5.717+5,-0.95-4) --(-4.517+5,-0.95-4);
\draw[scalar] (-4.517+5,-0.95-4)-- (-4.517+5,-2.15-4);
\draw[fermion] (-5.717+5,-2.15-4)-- (-4.517+5,-2.15-4);
\draw[fermion] (-4.517+5,-0.95-4) --(-3.54+5,-0.5-4) ;
\node at (-3.54+5,-0.2-4) {$ p'_{1}, a $};
\draw[fermion] (-4.517+5,-2.15-4) --(-3.54+5,-2.6-4) ;
\node at (-3.54+5,-2.9-4){$ p'_{2}, c $};
\node at (-5.117+5,-3.2-4){$6$};
\draw[fermion](-6.7+10,-2.6-4)  --(-5.717+10,-2.15-4);
\node at  (-6.7+10, -2.9-4){$ p_{2}, d $} ;
\draw[fermion](-6.7+10,-0.5-4) -- (-5.717+10,-0.95-4);
\node at (-6.7+10,-0.2-4) {$ p_{1}, b $};
\draw[scalar] (-5.717+10,-2.15-4) --(-5.717+10,-0.95-4);
\draw[fermion] (-5.717+10,-0.95-4) --(-4.517+10,-0.95-4);
\draw[decorate,decoration={snake,segment length=5,amplitude=1}] (-4.517+10,-0.95-4)-- (-4.517+10,-2.15-4);
\draw[fermion] (-5.717+10,-2.15-4)-- (-4.517+10,-2.15-4);
\draw[fermion] (-4.517+10,-0.95-4) --(-3.54+10,-0.5-4) ;
\node at (-3.54+10,-0.2-4){$ p'_{1}, a $};
\draw[fermion] (-4.517+10,-2.15-4) --(-3.54+10,-2.6-4) ;
\node at (-3.54+10,-2.9-4){$ p'_{2}, c $};
\node at (-5.117+10,-3.2-4){$7$};
\draw[fermion](-6.7+15,-2.6-4)  --(-5.717+15,-2.15-4);
\node at  (-6.7+15, -2.9-4){$ p_{2}, d $} ;
\draw[fermion](-6.7+15,-0.5-4) --(-5.717+15,-0.95-4);
\node at (-6.7+15,-0.2-4) {$ p_{1}, b $};
\draw[decorate,decoration={snake,segment length=5,amplitude=1}] (-5.717+15,-2.15-4)-- (-4.517+15,-0.95-4);
\draw[fermion] (-5.717+15,-0.95-4)--(-4.517+15,-0.95-4)  ;
\draw[draw=white,line width=4.5] (-5.4+15,-1.267-4) -- (-4.7+15,-1.967-4);
\draw[scalar] (-5.717+15,-0.95-4) -- (-4.517+15,-2.15-4);
\draw[fermion] (-5.717+15,-2.15-4)-- (-4.517+15,-2.15-4);
\draw[fermion] (-4.517+15,-0.95-4) --(-3.54+15,-0.5-4) ;
\node at (-3.54+15,-0.2-4) {$ p'_{1}, a $};
\draw[fermion] (-4.517+15,-2.15-4) --(-3.54+15,-2.6-4) ;
\node at (-3.54+15,-2.9-4){$ p'_{2}, c $};
\node at (-5.117+15,-3.2-4){$8$};
\draw[fermion](-6.7,-2.6-8)  --(-5.717,-2.15-8);
\node at  (-6.7, -2.9-8){$ p_{2}, d $} ;
\draw[fermion](-6.7,-0.5-8) --(-5.717,-0.95-8);
\node at (-6.7,-0.2-8) {$ p_{1}, b $};
\draw[scalar] (-5.717,-2.15-8)-- (-4.517,-0.95-8);
\draw[fermion] (-5.717,-0.95-8)--(-4.517,-0.95-8)  ;
\draw[draw=white,line width=4.5] (-5.4,-1.267-8) -- (-4.7,-1.967-8);
\draw[decorate,decoration={snake,segment length=5,amplitude=1}] (-5.717,-0.95-8) -- (-4.517,-2.15-8);
\draw[fermion] (-5.717,-2.15-8)-- (-4.517,-2.15-8);
\draw[fermion] (-4.517,-0.95-8) --(-3.54,-0.5-8) ;
\node at (-3.54,-0.2-8) {$ p'_{1}, a $};
\draw[fermion] (-4.517,-2.15-8) --(-3.54,-2.6-8) ;
\node at (-3.54,-2.9-8){$ p'_{2}, c $};
\node at (-5.117,-3.2-8){$9$};
\draw[fermion](-6.7+5,-2.6-8)  --(-5.117+5,-2.19952-8);
\node at  (-6.7+5, -2.9-8){$ p_{2}, d $} ;
\draw[fermion](-6.7+5,-0.5-8) -- (-5.867+5,-0.900481-8);
\node at (-6.7+5,-0.2-8) {$ p_{1}, b $};
\draw[decorate,decoration={snake,segment length=5,amplitude=1}] (-5.867+5,-0.900481-8)--(-4.367+5,-0.900481-8);
\draw[fermion] (-5.117+5,-2.19952+.5-8)--(-4.367+5,-0.900481-8);
\draw[fermion] (-5.867+5,-0.900481-8)--(-5.117+5,-2.19952+.5-8);
\draw[scalar] (-5.117+5,-2.19952+.5-8)-- (-5.117+5,-2.19952-8);
\draw[fermion] (-4.367+5,-0.900481-8) --(-3.54+5,-0.5-8) ;
\node at (-3.54+5,-0.2-8) {$  p'_{1},a $};
\draw[fermion] (-5.117+5,-2.19952-8) --(-3.54+5,-2.6-8) ;
\node at (-3.54+5,-2.9-8){$  p'_{2}, c $};
\node at (-5.117+5,-3.2-8){$10$};
\draw[fermion](-6.7+10,-2.6-8)  --(-5.867+10,-2.19952-8);
\node at  (-6.7+10, -2.9-8){$ p_{2},d  $} ;
\draw[fermion](-6.7+10,-0.5-8) -- (-5.117+10,-0.900481-8);
\node at (-6.7+10,-0.2-8) {$  p_{1}, b  $};
\draw[decorate,decoration={snake,segment length=5,amplitude=1}] (-5.867+10,-2.19952-8)--(-4.367+10,-2.19952-8);
\draw[scalar] (-5.117+10,-0.900481-8)--(-5.117+10,-0.900481-.5-8);
\draw[fermion] (-5.867+10,-2.19952-8) --(-5.117+10,-0.900481-.5-8);
\draw[fermion] (-5.117+10,-0.900481-.5-8)-- (-4.367+10,-2.19952-8);
\draw[fermion] (-5.117+10,-0.900481-8) --(-3.54+10,-0.5-8) ;
\node at (-3.54+10,-0.2-8) {$  p'_{1},a $};
\draw[fermion] (-4.367+10,-2.19952-8)  --(-3.54+10,-2.6-8) ;
\node at (-3.54+10,-2.9-8){$ p'_{2}, c  $};
\node at (-5.117+10,-3.2-8){$11$};
\draw[fermion](-6.7+15,-2.6-8)  --(-5.117+15, -1.55-0.7-8);
\node at  (-6.7+15, -2.9-8){$  p_{2},d $} ;
\draw[fermion](-6.7+15,-0.5-8) -- (-5.117+15, -1.55+0.7-8);
\node at (-6.7+15,-0.2-8) {$ p_{1}, b $};
\draw[fermion](-5.117+15, -1.55+0.4-8)  arc (90: -90:0.4);
\draw[fermion] (-5.117+15, -1.55-0.4-8) arc (270:90:0.4);
\draw[scalar] (-5.117+15, -1.55-0.4-8)--(-5.117+15, -1.55-0.7-8);
\draw[scalar] (-5.117+15, -1.55+0.4-8)--(-5.117+15, -1.55+0.7-8);
\draw[fermion](-5.117+15, -1.55+0.7-8) --(-3.54+15,-0.5-8) ;
\node at (-3.54+15,-0.2-8) {$  p'_{1},a$};
\draw[fermion] (-5.117+15, -1.55-0.7-8) --(-3.54+15,-2.6-8) ;
\node at (-3.54+15,-2.9-8){$  p'_{2}, c  $};
\node at (-5.117+15,-3.2-8){$12$};
\draw[fermion](-6.7,-2.6-12)  --(-5.717,-2.15-12);
\node at  (-6.7, -2.9-12){$ p_{2}, d $} ;
\draw[fermion](-6.7,-0.5-12) -- (-5.717,-0.95-12);
\node at (-6.7,-0.2-12) {$ p_{1}, b $};
\draw[scalar] (-5.717,-2.15-12) --(-5.717,-0.95-12);
\draw[fermion] (-5.717,-0.95-12) --(-4.517,-0.95-12);
\draw[scalar] (-4.517,-0.95-12)-- (-4.517,-2.15-12);
\draw[fermion] (-5.717,-2.15-12)-- (-4.517,-2.15-12);
\draw[fermion] (-4.517,-0.95-12) --(-3.54,-0.5-12) ;
\node at (-3.54,-0.2-12) {$ p'_{1}, a $};
\draw[fermion] (-4.517,-2.15-12) --(-3.54,-2.6-12) ;
\node at (-3.54,-2.9-12){$ p'_{2}, c $};
\node at (-5.117,-3.2-12){$13$};
\draw[fermion](-6.7+5,-2.6-12)  --(-5.717+5,-2.15-12);
\node at  (-6.7+5, -2.9-12){$ p_{2}, d $} ;
\draw[fermion](-6.7+5,-0.5-12) --(-5.717+5,-0.95-12);
\node at (-6.7+5,-0.2-12) {$ p_{1}, b $};
\draw[scalar] (-5.717+5,-2.15-12)-- (-4.517+5,-0.95-12);
\draw[fermion] (-5.717+5,-0.95-12)--(-4.517+5,-0.95-12)  ;
\draw[draw=white,line width=4.5] (-5.4+5,-1.267-12) -- (-4.7+5,-1.967-12);
\draw[scalar] (-5.717+5,-0.95-12) -- (-4.517+5,-2.15-12);
\draw[fermion] (-5.717+5,-2.15-12)-- (-4.517+5,-2.15-12);
\draw[fermion] (-4.517+5,-0.95-12) --(-3.54+5,-0.5-12) ;
\node at (-3.54+5,-0.2-12) {$ p'_{1}, a $};
\draw[fermion] (-4.517+5,-2.15-12) --(-3.54+5,-2.6-12) ;
\node at (-3.54+5,-2.9-12){$ p'_{2}, c $};
\node at (-5.117+5,-3.2-12){$14$};
\draw[fermion](-6.7+10,-2.6-12)  --(-5.117+10,-2.19952-12);
\node at  (-6.7+10, -2.9-12){$ p_{2}, d $} ;
\draw[fermion](-6.7+10,-0.5-12) -- (-5.867+10,-0.900481-12);
\node at (-6.7+10,-0.2-12) {$ p_{1}, b $};
\draw[scalar] (-5.867+10,-0.900481-12)--(-4.367+10,-0.900481-12);
\draw[fermion] (-5.117+10,-2.19952+.5-12)--(-4.367+10,-0.900481-12);
\draw[fermion] (-5.867+10,-0.900481-12)--(-5.117+10,-2.19952+.5-12);
\draw[scalar] (-5.117+10,-2.19952+.5-12)-- (-5.117+10,-2.19952-12);
\draw[fermion] (-4.367+10,-0.900481-12) --(-3.54+10,-0.5-12) ;
\node at (-3.54+10,-0.2-12) {$  p'_{1},a $};
\draw[fermion] (-5.117+10,-2.19952-12) --(-3.54+10,-2.6-12) ;
\node at (-3.54+10,-2.9-12){$  p'_{2}, c $};
\node at (-5.117+10,-3.2-12){$15$};
\draw[fermion](-6.7+15,-2.6-12)  --(-5.867+15,-2.19952-12);
\node at  (-6.7+15, -2.9-12){$ p_{2},d  $} ;
\draw[fermion](-6.7+15,-0.5-12) -- (-5.117+15,-0.900481-12);
\node at (-6.7+15,-0.2-12) {$  p_{1}, b  $};
\draw[scalar] (-5.867+15,-2.19952-12)--(-4.367+15,-2.19952-12);
\draw[scalar] (-5.117+15,-0.900481-12)--(-5.117+15,-0.900481-.5-12);
\draw[fermion] (-5.867+15,-2.19952-12) --(-5.117+15,-0.900481-.5-12);
\draw[fermion] (-5.117+15,-0.900481-.5-12)-- (-4.367+15,-2.19952-12);
\draw[fermion] (-5.117+15,-0.900481-12) --(-3.54+15,-0.5-12) ;
\node at (-3.54+15,-0.2-12) {$  p'_{1},a $};
\draw[fermion] (-4.367+15,-2.19952-12)  --(-3.54+15,-2.6-12) ;
\node at (-3.54+15,-2.9-12){$ p'_{2}, c  $};
\node at (-5.117+15,-3.2-12){$16$};
\end{tikzpicture}
    \caption{Feynman diagrams for the four-spinor vertex at one-loop level. All external momenta are incoming. There
are 16 additional diagrams obtained from the exchange $\{p'_1,a\}\leftrightarrow \{p'_2,c\}$.}
    \label{fig:ssss}
\end{figure}
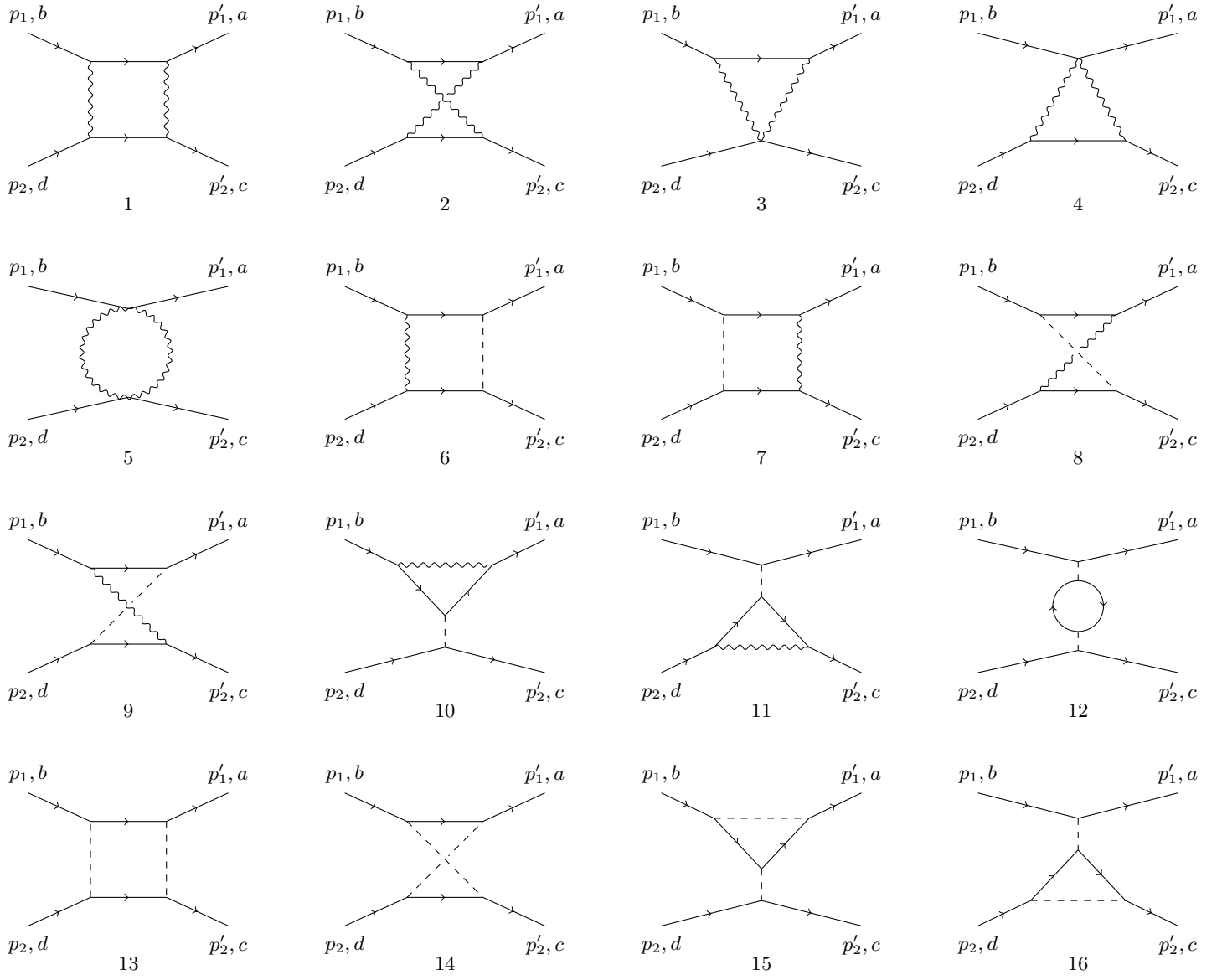
The last potentially divergent vertex allowed by the superficial degree of divergence is the four-spinor vertex. At one-loop level, this vertex receives contributions from the diagrams shown in Fig.~\ref{fig:ssss}. The complete one-loop correction can be written as
\begin{equation}
\label{4fr}
\begin{split}
i\Lambda_{abcd}\left(p_1,p_2,p'_1,p'_2\right)=&i\lambda_{abcd}
+i \Lambda^{*}_{abcd}\left(p_1,p_2,p'_1,p'_2\right) \\&+
i\left[\delta_{\lambda_1}\lambda_1
\mathds{1}_{ab}\mathds{1}_{cd}+\delta_{\lambda_{2}}\lambda_{2}\gamma^5_{ab}\gamma^5_{cd}+\vartheta\delta_{\lambda_{3}}
\lambda_{3} \gamma^{\mu}_{ab}\gamma_{\mu cd}\right]-\varsigma \left(\{p'_1,a\}\leftrightarrow \{p'_2,c\}\right),
\end{split}
\end{equation}
where $i \Lambda^{*}_{abcd}$ denotes the sum of the one-loop diagrams depicted in Fig.~\ref{fig:ssss}. 
We employ the modified minimal subtraction scheme ($\overline{\mathrm{MS}}$) and evaluate the ultraviolet-divergent part of the vertex in the zero-external-momentum limit. The corresponding counterterms are then fixed by requiring the cancellation of all one-loop ultraviolet divergences.
Therefore, the set of counterterms that yield a finite four-spinor vertex function at one-loop is
\begin{equation}
\label{deltalambda}
\begin{split}
\delta_{\lambda_1}=& -\frac{1}{(4\pi)^2\lambda_1}\frac{2}{\tilde{\epsilon}}
\Bigg\{3e^4\left(1-\frac{g^{2}}{4}\right)^2
+e^2\left\{\frac{g^2}{4}
\left[4(1-\vartheta)\lambda_1+(1-4\vartheta)\lambda_2-6\vartheta\lambda_3\right]+\xi \lambda_1\right\}\\&+\frac{(4-\tau)}{2}\lambda_1^2+\lambda_2(\lambda_1+\lambda_2)+4\vartheta\lambda_3(\lambda_1+\lambda_3)\Bigg\},\\
\delta_{\lambda_2}=&-\frac{2}{(4\pi)^2\lambda_2}\frac{1}{\tilde{\epsilon}} \left\{
e^2\left\{\frac{g^2}{4}
\left[(1-4\vartheta)\lambda_1+4(1-\vartheta)\lambda_2+6\vartheta\lambda_3\right]+\xi \lambda_2\right\}+\frac{(2-\tau)}{2}\lambda_2^2+3\lambda_1\lambda_2-4\vartheta\lambda_2\lambda_3\right\},
\\
\delta_{\lambda_3}=& -\frac{2\vartheta}{(4\pi)^2}\frac{1}{\tilde{\epsilon}} \left\{e^2\left(-\frac{3g^2}{4}
+\xi\right)+3 \lambda_{1}-\lambda_{2}-\frac{(4+\tau)}{2} \lambda_{3}\right\} .
\end{split}
\end{equation}  
With these counterterms, all ultraviolet-divergent contributions to the four-spinor vertex cancel, rendering the renormalized vertex function finite at one-loop order.

\subsection{Beta Functions}\label{subsec:beta}
Once the ultraviolet divergences have been absorbed into the corresponding counterterms, from the relation between the bare and renormalized
parameters of the theory:
\begin{equation}
\begin{array}{c}
\begin{array}{ccccc}
 e_0=Z_1^{-\frac{1}{2}}Z_2^{-1}Z_e\mu^{\epsilon} e ,&\quad & e_0^2=Z_1^{-1} Z_2^{-1} Z_3 \mu^{2\epsilon}e^2, &\quad&
\lambda_{0j}=Z_{2}^{-2}Z_{\lambda_j}\mu^{2\epsilon}\lambda_{j},
\end{array}
\\
\begin{array}{ccc}
g_0=Z_e^{-1}Z_{eg}g,&\quad& m_0^2=Z_2^{-1}Z_mm^2,
\end{array}
\end{array}
\label{ctc3}
\end{equation}
and the renormalization constants written in the $\overline{\text{MS}}$ scheme:
\begin{eqnarray}
&&Z^{\overline{\text{MS}}}_1=1 - \frac{  e^2\tau } { (4\pi)^2}\left( \frac{g^2}{4}-\frac{1}{3}\right)
\frac{1}{\tilde\epsilon},\label{ctc4}\\
&&Z^{\overline{\text{MS}}}_2=Z^{\overline{\text{MS}}}_3=
Z^{\overline{\text{MS}}}_e=1+\frac{e^2}{(4\pi)^2}\left(3-\xi\right)\frac{1}{\tilde\epsilon},\\
&&Z^{\overline{\text{MS}}}_{\lambda_1}= 1-\frac{1}{(4\pi)^2\lambda_1}\frac{2}{\tilde{\epsilon}}
\Bigg\{3e^4\left(1-\frac{g^{2}}{4}\right)^2
+e^2\left\{\frac{g^2}{4}
\left[4(1-\vartheta)\lambda_1+(1-4\vartheta)\lambda_2-6\vartheta\lambda_3\right]+\xi \lambda_1\right\}\\&&\qquad\qquad+\frac{(4-\tau)}{2}\lambda_1^2+\lambda_2(\lambda_1+\lambda_2)+4\vartheta\lambda_3(\lambda_1+\lambda_3)\Bigg\},\nonumber\\
&&Z^{\overline{\text{MS}}}_{\lambda_2}=1-\frac{2}{(4\pi)^2\lambda_2}\frac{1}{\tilde{\epsilon}} \Bigg\{
e^2\left\{\frac{g^2}{4}
\left[(1-4\vartheta)\lambda_1+4(1-\vartheta)\lambda_2+6\vartheta\lambda_3\right]+\xi \lambda_2\right\}\\&&\qquad\qquad+\frac{(2-\tau)}{2}\lambda_2^2+3\lambda_1\lambda_2-4\vartheta\lambda_2\lambda_3\Bigg\},\nonumber\\
&&Z^{\overline{\text{MS}}}_{\lambda_3}=1 -\frac{2\vartheta}{(4\pi)^2}\frac{1}{\tilde{\epsilon}} \left\{e^2\left(-\frac{3g^2}{4}
+\xi\right)+3 \lambda_{1}-\lambda_{2}-\frac{(4+\tau)}{2} \lambda_{3}\right\},\nonumber\\
&&Z^{\overline{\text{MS}}}_{eg}=Z^{\overline{\text{MS}}}_{e}+\delta^{\overline{\text{MS}}}_{g} =
1+ \frac{1}{(4\pi)^2} 
\frac{1}{\tilde{\epsilon}}\left[\left(\frac{g^{2}}{4}+2-\xi\right)e^2
 - \lambda_1-\lambda_{2}\right],\nonumber\\
&& Z^{\overline{\text{MS}}}_{m}=Z^{\overline{\text{MS}}}_{2}+\delta^{\overline{\text{MS}}}_m = 1 -\frac{1}{(4\pi)^2}\frac{1}{\tilde{\epsilon}}
\left[e^2\left(\frac{3g^2}{4}+\xi\right)+(1-\tau)\lambda_1+\lambda_{2}
+4\vartheta\lambda_{3}  \right] 
,\label{ctc4b}
\end{eqnarray}
one can extract the beta functions $\beta_\eta\equiv\mu\frac{\partial\eta}   {\partial\mu}$ and anomalous dimensions $\gamma_m\equiv\frac{\mu}{m}\frac{\partial m}{\partial\mu}$ in the $\epsilon\to 0$ limit. This yields
\begin{eqnarray}
&&\beta_e=\frac{e^3\tau}{48\pi^2}\left(\frac{3}{4}g^2-1\right),\label{betafu1}\\
&&\beta_g=\frac{g}{8\pi ^2}\left[e^2 \left(\frac{g^2}{4}-1\right)-\lambda_{1}-\lambda_{2}\right],\\
&& \beta_{\lambda_1}=-\frac{1}{16\pi^2}\Bigg\{\frac{3}{4} e^4
\left(g^2-4\right)^2+ e^2 \big\{ g^2\left[4(1-\vartheta)\lambda_1+(1-4\vartheta)\lambda_2-6\vartheta\lambda_3\right]
+12\lambda_{1}\big\}\label{eq:betalambda1}\\&&\qquad\qquad +2(4-\tau)\lambda_1^2+4\lambda_2\left(\lambda_1+\lambda_2\right)+16\vartheta\lambda_3\left(\lambda_1+\lambda_3\right)\Bigg\},\nonumber\\
&&\beta_{\lambda_2}=-\frac{1}{16 \pi
   ^2} \Big\{ e^2 \big\{g^2\left[(1-4\vartheta)\lambda_1+4(1-\vartheta)\lambda_2+6\vartheta\lambda_3\right]+12\lambda_2\big\}\\&&\qquad\qquad+2\lambda_2\left[6\lambda_1+(\tau-2)\lambda_2-8\vartheta\lambda_3\right]\Big\},\nonumber\\
&&\beta_{\lambda_3}=-\frac{\vartheta\lambda_3}{16 \pi ^2}\Big\{3e^2 (4-g^2)+2(6\lambda_1-2\lambda_2-(4+\tau)\lambda_3)\Big\},\nonumber\\
&&\gamma_m=-\frac{1}{16 \pi ^2}\left\{3 e^2 \left(\frac{g^2}{4}+1\right)+(1-\tau) \lambda_{1}+ \lambda_{2}+4\vartheta \lambda_{3}\right\}.\label{betafu2}
\end{eqnarray}
As expected, the beta functions are independent of the choice of the gauge-fixing parameter $\xi$, and therefore gauge-invariant.
Finally, to obtain the Renormalization Group Equations (RGE) corresponding to F-SOPHY and B-SOPHY, we apply the respective conditions on $\tau=4\varsigma$ and $\vartheta=\Theta(\varsigma)$.
Therefore, for F-SOPHY we have 
\begin{eqnarray}
&&\beta^{F}_e=\frac{e^3}{12\pi^2}\left(\frac{3}{4}g^2-1\right),\label{betafuf1}\\
&&\beta^{F}_g=\frac{g}{8\pi ^2}\left[e^2 \left(\frac{g^2}{4}-1\right)-\lambda_{1}-\lambda_{2}\right],\\
&& \beta^{F}_{\lambda_1}=-\frac{1}{4\pi^2}\Bigg\{3 e^4
\left(\frac{g^2}{4}-1\right)^2-3e^2 \left[ \frac{g^2}{4}\left(\lambda_2+2\lambda_3\right)
-\lambda_{1}\right]+\lambda_2\left(\lambda_1+\lambda_2\right)+4\lambda_3\left(\lambda_1+\lambda_3\right)\Bigg\},\\
&&\beta^{F}_{\lambda_2}=\frac{1}{4 \pi
   ^2} \Bigg\{ 3e^2 \left[\frac{g^2}{4}\left(\lambda_1-2\lambda_3\right)-\lambda_2\right]-\lambda_2\left[3\lambda_1-\lambda_2-4\lambda_3\right]\Bigg\},\\
&&\beta^{F}_{\lambda_3}=-\frac{\lambda_3}{4 \pi ^2}\Bigg\{3e^2 \left(1-\frac{g^2}{4}\right)+3\lambda_1-\lambda_2-4\lambda_3\Bigg\},\\
&&\gamma^{F}_m=-\frac{1}{16 \pi ^2}\left\{3 e^2 \left(\frac{g^2}{4}+1\right)-3\lambda_{1}+ \lambda_{2}+4 \lambda_{3}\right\},\label{betafuf2}
\end{eqnarray}
while, for B-SOPHY
\begin{eqnarray}
&&\beta^{B}_e=-\frac{e^3}{12\pi^2}\left(\frac{3}{4}g^2-1\right),\label{betafub1}\\
&&\beta^{B}_g=\frac{g}{8\pi ^2}\left[e^2 \left(\frac{g^2}{4}-1\right)-\lambda_{1}-\lambda_{2}\right],\\
&& \beta^{B}_{\lambda_1}=-\frac{1}{4\pi^2}\Bigg\{3e^4
\left(\frac{g^2}{4}-1\right)^2+ e^2 \left[ \frac{g^2}{4}\left(4\lambda_1+\lambda_2\right)
+3\lambda_{1}\right]+4\lambda_1^2+\lambda_2\left(\lambda_1+\lambda_2\right)\Bigg\},\\
&&\beta^{B}_{\lambda_2}=-\frac{1}{4 \pi
   ^2} \Bigg\{ e^2 \left[\frac{g^2}{4}\left(\lambda_1+4\lambda_2\right)+3\lambda_2\right]+3\lambda_2\left(\lambda_1+\lambda_2\right)\Bigg\},\\
&&\gamma^{B}_m=-\frac{1}{16 \pi ^2}\left\{3 e^2 \left(\frac{g^2}{4}+1\right)+5\lambda_{1}+ \lambda_{2}\right\}.\label{betafub2}
\end{eqnarray}
Notice that for both cases, $g$ and $\lambda_i$, $i=1,2,3$, have a fixed point at $g=\pm2$ for vanishing self-interactions, and $g$ has another fixed point at $g=0$. Moreover, for F-SOPHY, $\beta^{F}_e$ and $\gamma^{F}_m$ evaluated at $g=\pm2$, $\lambda_i=0$ differ from their Dirac QED counterparts
\begin{eqnarray}\label{eq:betaDir}
\beta^D_e=\frac{e^3}{12\pi^2},\qquad
\gamma^D_m=-\frac{3 e^2}{8 \pi ^2},
\end{eqnarray}
by a factor of $2$ in $\beta^{F}_e$, this is consistent with the fact that F-SOPHY has twice the number of degrees of freedom with respect to the Dirac formalism. This correspondence has been known for a long time and constitutes the first application of F-SOPHY: Calculations performed at this fixed point of the theory reproduce the corresponding  Dirac results once the difference in the number of degrees of freedom is properly adjusted~\cite{Morgan:1995te,Veltman:1997am}. It is straightforward to show that indeed Eq.~\eqref{eq:betaDir} can be obtained from Eqs.~(\ref{betafu1}-\ref{betafu2}), by setting $g=2,\lambda_i=0, \vartheta=1$ and artificially taking $\tau= 2$. For $g\neq \pm 2$,
the scalar-scalar self-interaction parameterized by  $\lambda_{1}$ is generated at one loop level, even for vanishing self-interactions at tree-level, while  $\lambda_2$ has a fixed point at $\lambda_i=0$, and $\lambda_3$ at $\lambda_3=0$. 

Another familiar result emerges from B-SOPHY in the other fixed point for the Pauli coupling $g=0$. At this point, the spin is decoupled from the gauge field, and the spin degrees of freedom just collapse into a multiplicative factor. This theory resembles a collection of scalar KG fields but is enhanced with a pseudoscalar-pseudoscalar interaction.  In fact, one can even artificially recover the familiar scalar QED beta functions by taking $g=0$, $\lambda_2=0$, $\vartheta=0$, $\tau=-1$ and $\lambda_1\to-\frac{\lambda_\phi}{2}$ in  Eqs.~(\ref{betafu1}-\ref{betafu2}), yielding 
\begin{eqnarray}
&&\beta^S_e=\frac{e^3}{48\pi^2},\label{betaScal}\\
&&\beta^S_{\lambda_\phi}=-2\beta^S_{\lambda_1}|_{\lambda_1\to-\lambda_\phi/2}=\frac{1}{16 \pi
   ^2} \left( 24e^4-12e^2\lambda_{\phi}+5\lambda_{\phi}^2\right),\nonumber\\
&&\gamma^S_m=-\frac{1}{16 \pi ^2}\left(3 e^2 -\lambda_{\phi}\right).\nonumber
\end{eqnarray}

It is worth noting that F-SOPHY with $g^2<4/3$ and B-SOPHY with $g^2>4/3$ have $\beta_e<0$ at one-loop level, suggesting that these $U(1)$ gauge theories in four dimensions may display asymptotic freedom without invoking a non-Abelian gauge theory, and therefore evading the Coleman--Gross theorem~\cite{Coleman:1973sx}.   A detailed analysis will be performed in a future paper. Finally, another attractive feature of the theory is the case $e=0$, $eg=0$, where the gauge and Pauli interactions are turned off. In this limit,
we are left with a renormalizable theory of self-interacting fermions, analogous to the one originally proposed by Nambu and Jona-Lasinio~\cite{Nambu:1961tp, Nambu:1961tp}, with a rich set of RGEs.

\subsection{Self-interactions of self-conjugate fields}
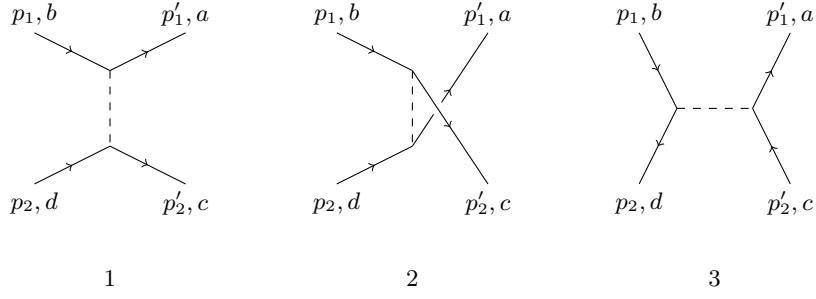
\begin{figure}
    \centering
\begin{tikzpicture}[scale=1.0]
\draw[fermion] (-6,-7+.5)--(-5,-7.5+.5);
\draw[fermion] (-5,-7.5+.5)--(-4,-7+.5);
\draw[fermion] (-6,-9+.5)  --(-5,-8.5+.5) ;
\draw[fermion] (-5,-8.5+.5)  --  (-4,-9+.5);
\draw[scalar] (-5,-7.5+.5) --(-5,-8.5+.5);
\node at (-6,-9.25+.5){$p_{2},d$};
\node at (-4,-9.25+.5) {$p^{\prime}_{2},c$};
\node at (-6,-6.75+.5) {$p_{1},b$};
\node at (-4,-6.75+.5) {$p^{\prime}_{1},a$};
\node at (-5,-10+.25) {$1$};
\draw[fermion] (-6+4,-7+.5)--(-5+4,-7.5+.5);
\draw[fermion] (-6+4,-9+.5)  --(-5+4,-8.5+.5) ;
\draw[fermion] (-5+4,-8.5+.5)  --  (-4+4,-7+.5);
\draw[draw=white,line width=5] (-5+4,-7.5+.5)--(-4+4,-9+.5);
\draw[fermion] (-5+4,-7.5+.5)--(-4+4,-9+.5);
\draw[scalar] (-5+4,-7.5+.5) --(-5+4,-8.5+.5);
\node at (-6+4,-9.25+.5){$p_{2},d$};
\node at (-4+4,-9.25+.5) {$p^{\prime}_{2},c$};
\node at (-6+4,-6.75+.5) {$p_{1},b$};
\node at (-4+4,-6.75+.5) {$p^{\prime}_{1},a$};
\node at (-5+4,-10+.25) {$2$};
\draw[fermion] (2,-6.5)--(2.5,-7.5);
\draw[fermion] (2.5,-7.5)--(2,-8.5);
\draw[scalar] (2.5,-7.5) --(3.5,-7.5);
\draw[fermion] (3.5,-7.5)--(4,-6.5);
\draw[fermion] (4,-8.5)--(3.5,-7.5);
\node at (-6+8,-9.25+.5){$p_{2},d$};
\node at (-4+8,-9.25+.5) {$p^{\prime}_{2},c$};
\node at (-6+8,-6.75+.5) {$p_{1},b$};
\node at (-4+8,-6.75+.5) {$p^{\prime}_{1},a$};
\node at (-5+8,-10+.25) {$3$};
\end{tikzpicture}
    \caption{ Feynman diagrams contributing to the tree-level four-spinor vertex function for M-SOPHY.}
    \label{fig:selfM}
\end{figure}

According to the analysis presented in Section \ref{sec:Fierz}, M-SOPHY admits only one independent interaction
\begin{equation}
\begin{split}
\mathcal{L}_M= \frac{1}{2}\partial^{\mu}\widehat{\psi}_M\partial_\mu\psi_M-\frac{m^2}{2}\widehat{\psi}_M\psi_M+\frac{\lambda_1}{12}\left(
\widehat{\psi}_M
\psi_M
\right)^2.
\end{split}\label{self_conjugate_int}
\end{equation}
The crucial difference from the previous theories is that the self-interaction contribution to the tree-level
four-spinor function is given by 
the three diagrams of Fig.~\ref{fig:selfM}, and yields $i(\lambda_{abcd}-\lambda_{cbad}+\lambda_{acdb})$.\\
The one-loop renormalization of this theory involves diagrams 3 and 4 of Fig.~\ref{fig:self}, and diagrams 12 to 16 of Fig.~\ref{fig:ssss}, together with the 10 additional diagrams obtained from the exchanges $\{p'_1,a\}\leftrightarrow \{p'_2,c\}$ and $\{p'_1,a\}\leftrightarrow \{p_2,d\}$. The corresponding RGEs actually coincide with Eqs.~\eqref{eq:betalambda1} and \eqref{betafu2} evaluated at $e=0$, $g=0$, $\lambda_2=\lambda_3=0$, $\vartheta=1$ and taking $\tau=2$, since in this case every closed fermion loop involves the exchange of two identical particles, and therefore a symmetry factor $(1/2)$ must be attached to each trace, consistent with the fact that the self-conjugate field carries only four degrees of freedom. Thus, the beta function and anomalous dimension for this theory are given by
\begin{equation}
\beta_{\lambda_1}=-\frac{\lambda_1^2}{4\pi^2},\qquad\gamma_m=\frac{\lambda_{1}}{16 \pi ^2}.\label{betaself}
\end{equation}
This remarkably simple pseudo-Hermitian theory has a negative one-loop beta function and therefore may develop asymptotic freedom without even invoking a gauge theory. In a future paper, the two-loop renormalization of this model will be addressed in order to reach a definite conclusion.

\section{Conclusions}\label{sec:con}
In this work, we have built from first principles the three different incarnations of spin-$1/2$ SOPHIES, namely: Fermionic, Bosonic, and Fermionic Majorana-like. We have outlined the canonical quantization and the symmetries of the three theories. In contrast to the 
Dirac formalism, our second-order formalism features fields of mass dimension one. Thus, it has a richer set of potentially renormalizable interactions, including  quartic self-interactions. We have identified the set of independent self-interactions for each SOPHY and conclude that there are three for F-SOPHY, two for B-SOPHY and only one for M-SOPHY. 

We have studied the one-loop renormalization for F- and B-SOPHIES, deriving the RGEs for three different interactions: Minimal coupling with a $U(1)$ gauge field, non-minimal coupling with the same gauge field through a Pauli term (a gyromagnetic-like coupling) and four-spinor self-interactions. In the case of M-SOPHY, we have studied the one-loop renormalization of its only quartic self-interaction. All interactions admitted by each SOPHY have been shown to be renormalizable at one-loop level in an arbitrary covariant gauge using dimensional regularization  and the naive prescription for the chirality operator, which does not lead to inconsistencies since $\gamma^5$ appears an even number of times inside loops for all  processes involved. 

For F- and B-SOPHIES we have found that one-loop renormalization leads to a rich set of RGEs with deep connections to known theories, like the Dirac formalism and the KG scalar through the fixed points of the Pauli coupling $g=2$ and $g=0$, respectively. In addition, F-SOPHY with $g^2<4/3$ and B-SOPHY with $g^2>4/3$ have $\beta_e<0$ at one-loop level, suggesting that these $U(1)$ gauge theories in four dimensions may display asymptotic freedom without invoking a non-Abelian gauge theory and, therefore, evading the Coleman--Gross theorem. A more striking result is the fact that the only quartic coupling in M-SOPHY has a negative one-loop beta function and therefore may be
asymptotically free without the need for a gauge theory. A more detailed analysis of these cases will be presented in a future paper.

\section*{Acknowledgements}
 \noindent{\bf I.D.S.} is supported by  Secretar\'ia de Ciencia, Humanidades, Tecnolog\'ia e Innovaci\'on (Secihti) through the programs {\it``Estancias Posdoctorales por México'' and ``Sistema Nacional de Investigadoras e Investigadores (SNII)''}. {\bf A. G. R.} acknowledges support from Secihti (CBF-2026-310), SNII, and PROFEXCE (México).
 \noindent{\bf A.C.R.-P.} is supported by Secihti through the program {\it``Becas Nacionales''}. 
 {\bf C.A.V.-A.} is supported by Secihti through the program {\it``Investigadoras e Investigadores por M\'exico''} project 749 and SNII 58928.
\bibliography{bibliography}

\end{document}